\documentclass[12pt,a4paper]{article}
\usepackage[utf8]{inputenc}
\usepackage{authblk}
\usepackage{graphicx}
\usepackage{float}
\usepackage{color}
\usepackage{xcolor}
\usepackage{caption}
\usepackage{subcaption}
\usepackage{makeidx}
\usepackage{array}
\usepackage[T1]{fontenc}
\usepackage[utf8]{inputenc}
\usepackage{cite}
\usepackage[font=small,labelfont=bf]{caption}
\usepackage{graphicx}
\newcolumntype{P}[1]{>{\centering\arraybackslash}p{#1}}
\newcolumntype{M}[1]{>{\centering\arraybackslash}m{#1}}
\author[]{A. Arul Anne Elden}
\author[]{M. Ponmurugan {\footnote{email correspondence: ponphy@cutn.ac.in}}}
\affil[]{School of Basic and Applied Sciences, Central University of Tamil Nadu, \\Thiruvarur - 610 005, Tamil Nadu, India.}
\date{}
\usepackage[left=2cm,right=2cm,top=2cm,bottom=2cm]{geometry}

\begin{document}
\title{Single-walled Ising nanotube with opposite sign of interactions using Wang-Landau algorithm}
\maketitle
\section*{Abstract} 
The effect of opposite sign of interactions in a single-walled Ising nanotube is investigated using the Wang-Landau algorithm. The thermodynamic observables are calculated from the estimated density of states (DOS) with and without the presence of an external magnetic field. Irrespective of the applied magnetic field, a symmetric trend of DOS is observed for opposite sign of interactions which is in contrast to the asymmetric trend for same sign of interactions. Further, two types of anti-ferromagnetic (AFM) orderings, namely A-type and C-type anti-ferromagnetic order, are observed for opposite sign of interactions. These AFM spin orientations are switched to ferromagnetic (FM) phase by increasing the applied  magnetic field ($B$). However, the spin ordering changes from the ordered AFM/FM phase to a disordered paramagnetic phase by increasing the temperature. Phase diagram shows that these three phases coexist around $B=2.0$. This study indicates that, by properly tuning the magnetic properties, the single-walled nanotube can be used for fabrication of new types of magnetic storage nano materials.
\newline
Key words: Single-walled Ising nanotube, opposite sign of interactions, Wang-Landau algorithm, and phase transition.

\section{Introduction}
$~~~~$ In recent decades, the low dimensional magnetic systems has been intensively studied due to their exotic magnetic properties and vast technological applications including biomedicines \cite{Pankhurst2003Applications}, biomolecular motors \cite{Soong2000Powering}, spintronics devices \cite{Koch1998Magnetization}, energy storage \cite{Xie2020Hierarchical}, magneto-optic recording \cite{Shieh1986Magneto}, and nanocomposite magnets \cite{Zeng2002Exchange}. The magnetic nanosystems has various structures viz, nanoparticles, nanotubes, nanowires, nanoribben, nanographene, etc,. Many experimental techniques are used to synthesis nanomaterials like, CoNi nanotubes \cite{Zhang2013Fabri}, FeNi nanotubes \cite {Zhou2007Temp}, and carbon nanotubes \cite{Krainoi2022Novel}. Theoretical studies also play a vital role in understanding the magnetic nanosystems via different techniques including Green function technique \cite{Mi2016Magnon}, differential operator technique \cite{Farchakh2020Magneti}, mean field approximation \cite{Deviren2022Dynamic}, effective field theory \cite{Canko2014Hysteresis} and Monte Carlo technique \cite{Masrour2020Monte}. Among the interacting models available in the literature, the Ising model is one of the simplest model to study the phase transition in magnetic systems. The main advantage of this model is that any interacting system can be easily mapped to this model \cite{Wolf2000The, Louwerse2022Multi}, and it is more convenient to study the physical properties of low dimensional systems.

Several studies are reported in the context of Ising modelled nanosystems. In particular, Mn doped ZnO nanotubes were investigated using the Monte Carlo techniques, and analyzed for the effect of Mn concentration on magnetic properties \cite{Jabar2017Magnetic}. Some of the hexagonal Ising nanotubes with core-shell structures were reported to have ferrimagnetic mixed spins \cite{Hachem2021Phase}. Core-shell nanotube was also studied using metropolis algorithm to analyze the compensation temperature, magnetic properties, and phase diagram \cite{Wang2018Compen}. The pseudocritical magnetic properties for a single-walled ferromagnetic nanotube with varying system size and varying topology were reported \cite{Salazar2012Influence}. Some studies explored the magnetic properties of hexagonal Ising nanotube by applying the effective field theory coupled with differential operator technique. These techniques show that some systems with opposite sign of exchange interaction suffer magnetic frustration \cite{ElMaddahi2019Magnetic, Maddahi2017Magnetic, ElMaddahi2020Magnetic}. Our interest is to study the magnetic properties and the phase behavior of single walled Ising nanotube (SWINT) influenced by the  opposite sign of interaction between the inter-layer and the intra-layer using the Monte Carlo technique.

Apart from  the studies of ferromagnetic interaction, several studies of  Anti Ferro Magnetic (AFM) interactions in the layered lattice systems were also investigated and found to have different types of ground state \cite{Badiev2022Ground, Murtazaev2020Phase}. The AFM interactions in the alternative layers of the Ising system exhibit phase changes \cite{Chandra2020Monte, Acharyya2022Monte}. The experimental and DFT studies of the magnetic materials exhibit different types of magnetic spin orientations. The A-type AFM order was observed for the layered  $EuMg_2Sb_2$ single crystals \cite{Pakhira2022A-type, Marshall2021Magnetic}, its crystallographic, magnetic, and electronic properties are indicating a narrow gap semiconducting property.

In the case of nanotube system, Salazar et al. have studied the magnetic properties of ferromagnetic single walled nanotube using metropolis algorithm \cite{Salazar2012Monte, Salazar2012Monte1}. However, there are few simulation studies of AFM ordering for SWINT \cite{Elden2022Monte}. In our previous study, the ferromagnetic (FM) and different types of AFM orderings were observed for SWINT. The metropolis algorithm is used to examine the system and to analyze its phase transition \cite{Elden2022Monte}. We have also observed the asymmetric trend of Density of States (DOS) of SWINT under the applied magnetic field with the same sign of interactions using Wang-Landau (WL) algorithm. The WL algorithm has been widely used among the non-Boltzmann sampling techniques \cite{Landau2000A_guide}. The advantages of WL algorithm is that one can estimate DOS and then calculate the average thermodynamics properties at any finite temperature. The main focus of the present study is to investigate the effect of opposite sign of interactions in a SWINT with the absence/presence of a external magnetic field by employing WL algorithm.

The manuscript is presented as follows: Section 2 describes the SWINT model and WL algorithm. The discussion on the observed results with and without magnetic field is provided in the Section 3. Finally, the Conclusion section summarizes the important findings.

\section{Model and Simulation method} 
$~~~~$The spin-half Ising model is tailored as a single-walled Ising nanotube (SWINT) with a layered graphite-like structure \cite{Elden2022Monte}, as shown in Fig.\ref{f1}. Each layer comprise of six lattice sites with magnetic moments $S_i$ with values of $+1$ for spin up and $-1$ for spin down. The Hamiltonian of the system is,
\begin{equation}
\label{eq:1}
H ~=~-J_1\sum_{<i,j>} S_i S_j ~- J_2\sum_{<i,k>} S_i S_k ~- B\sum_{i} S_i,
\end{equation}
where, $J_1$ denotes strength of the intra-layer interaction, $J_2$ is strength of the inter-layer interaction, $< \cdots >$ represents the nearest neighbour spin sites, and $B$ is the external magnetic field. The system exhibits FM (parallel) spin ordering only if both the interactions $J_1$ and $J_2$ are greater than zero. The different types of AFM (anti-parallel) spin orderings are formed from the other combinations of signs of $J_1$ and $J_2$. Depending on such combination of interactions, the system is classified into two kinds: same sign ($J_1>0$; $J_2>0$, and $J_1<0$; $J_2<0$) and opposite sign of interactions ($J_1>0$; $J_2<0$, and $J_1<0$; $J_2>0$). 
Since some systems with opposite sign of exchange interaction suffers magnetic frustration \cite{ElMaddahi2019Magnetic, Maddahi2017Magnetic, ElMaddahi2020Magnetic}, this work will pay special attention to the opposite sign of interactions of $J_1$ and $J_2$ in SWINT, which arranges spins in A-type and C-type AFM orderings. In A-type, the spin alignment is parallel within the layer and anti-parallel between the layer. In contrast, C-type AFM order is observed by anti-parallel alignment of spins within the layer and parallel alignment of spins between the layers \cite{Wqllan1955Neutron}. The representation of these orderings are given in Fig. \ref{f1}. The closed boundary condition is applied for the $x$ and $y$ axes, whereas the periodic boundary condition is applied for the $z$ axis. The Wang-Landau algorithm were applied to the SWINT system to compute the density of states
($g(E)$). It is a two step process: first step is the estimation of the DOS and second step is the calculation of average observables from production run.

The algorithm performs a random walk in an energy space with  probability $P(E)\propto \frac{1}{g(E)}$ \cite{Wang2001Efficient, Wang2001Determining, Taylor2020Effects}. The algorithm begins with the initialization of DOS, $\ln g(E)=0$; histogram $H(E)=0$, and a modification factor $\ln f$. The initial configuration $C_i$ and the trail configuration $C_t$ are generated with energy $E_i$ and $E_t$, respectively. The transition (from $C_i$ to $C_t$) is accepted with transition probability,
\begin{eqnarray}
	\label{eq:5}
	P(C_i\rightarrow C_t)&=&min\Bigg(1,\Big[\ln g(E_i)-\ln g(E_t)\Big] \Bigg).
\end{eqnarray}
Further, the DOS and the histogram have been updated as $\ln~[g(E)] = \ln~[g(E)] + \ln f$ and $H(E)= H(E)+1$, respectively. If the histogram's flatness reaches a certain level ($80\%$), the modification factor is updated to $\ln f =\ln f\times 0.5$, and the histogram is then reset to $H(E)=0$. The algorithm gets terminated when the $\ln f$ reaches a small enough value ($\approx 10^{-8}$). Finally, we obtain a converged DOS from this algorithm.

An entropic ensemble is generated during the production run, and the canonical ensemble averages 
of any quantities are calculated at a given finite temperature as \cite{Murthy2000Monte, Jayasri2005Wang},
\begin{eqnarray}
	\label{eq:11}
	\langle O \rangle_\beta &=& \frac{\sum\limits_{C} O(C)~ g(E(C))~\exp[-\beta E(C)]}{\sum\limits_{C} g(E(C))~\exp[-\beta E(C)]},
\end{eqnarray}
where $C$ represents the configuration, $\beta=\frac{1}{k_ B T}$ ($k_B$ is a Boltzmann constant, taken as $1$ for entire simulation), $T$ is the temperature.  It is to be noted that all the parameters used in this simulation has an appropriate units in terms of $k_BT$ and magnetic field. The system's average magnetization, $\langle M \rangle$ and average energy, $\langle E \rangle$ are calculated using equation (\ref{eq:11}). The magnetization per spin is given by $M = \frac{1}{N}\sum_{i} S_i$. Considering there are $L$ number of layers, the total number of spins can be $N=6L$ as there are $6$ spins in each layer. The whole simulation is performed with $180$ spins in $30$ layers. The magnetic susceptibility is calculated as,
\begin{eqnarray}
\label{eq:4} 
\chi(T)&=&\frac{1}{k_{B}T} ~(\langle M^2 \rangle_T - \langle M \rangle^2_T).
\end{eqnarray}
The free energy $F(T)=-k_{B}T\ln(Z)$, and entropy $S(T)=\frac{U(T)-F(T)}{T}$ are also calculated, where $Z=\sum\limits_{E}g(E)~\exp[-\beta E]$ is the partition function and $U(T)=\langle E \rangle_T$ is the internal energy \cite{Suman2019Non-}. 
The following section analyze the effect of opposite sign of interactions in a SWINT subjected to the external external magnetic field by employing WL algorithm.

\section{Simulation results and discussions}
\begin{figure}[H]			
	\centering
	\begin{subfigure}[b]{0.2\textwidth}
		\includegraphics[width=\textwidth]{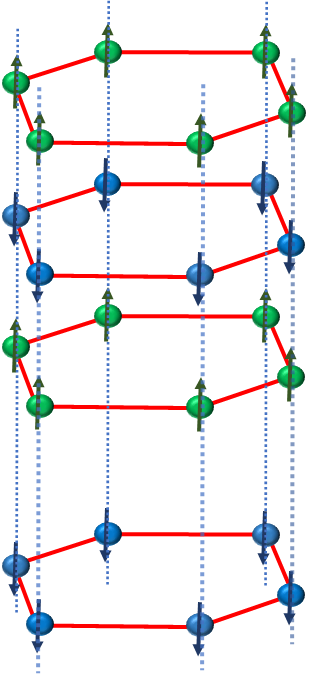}
		\caption{$ $}
		\label{f1a}
	\end{subfigure}
~
	\begin{subfigure}[b]{0.2\textwidth}
		\includegraphics[width=\textwidth]{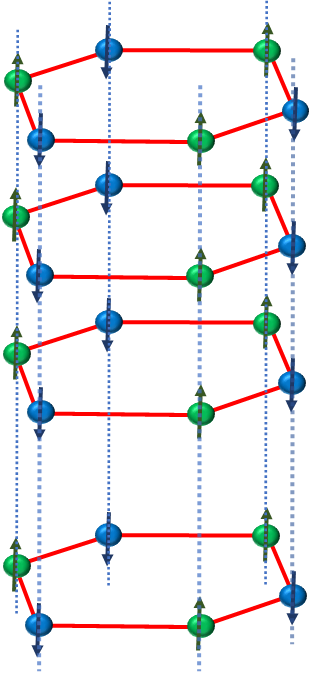}
		\caption{$ $}
		\label{f1b}
	\end{subfigure}
~
	\begin{subfigure}[b]{0.15\textwidth}
		\includegraphics[width=\textwidth]{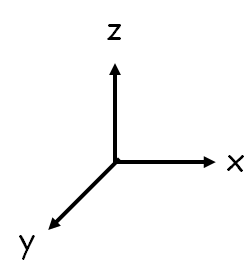}
	\end{subfigure}
~
	\caption{Schematic representation of the SWINT. The ground state spin orientation of opposite sign of interactions at $B=0$, (a) A-type AFM: $J_1=+1; ~J_2=-1$, and (b) C-type AFM:  $J_1=-1; ~J_2=+1$.}
\label{f1}
\end{figure}

$~~~~$The ground state spin ordering is obtained in the absence of a magnetic field for opposite sign of interactions, depicted in Fig. \ref{f1}. The A-type AFM ordering is obtained by applying the following interactions $J_1 =+1$ and $J_2 =-1$ (Fig. \ref{f1a}). The spins follow parallel ordering within the layer, and anti-parallel ordering between the layers.  The interactions $J_1=-1$ and $J_2 =+1$ are applied to the system, yielding the C-type AFM (Fig. \ref{f1b}). This ordering follow anti-parallel orientation within the layer, and a parallel orientation between the layers. So in this type of ordering, the spins are aligned similarly among all the layers.

In the entire simulation, the magnitude of the interaction $J_2$ kept fixed as ($J_2=|\pm 1|$) and its sign fixed depending on the ordering. Initially the system is investigated in the absence of magnetic field by varying the parameter $J_1$, and later by varying the magnetic field $B$. The simulation results are discussed in the proceeding section.

\subsection{In the absence of magnetic field}
\begin{figure}[H]
	\centering
	\begin{subfigure}[b]{0.4\textwidth}
		\includegraphics[width=\textwidth]{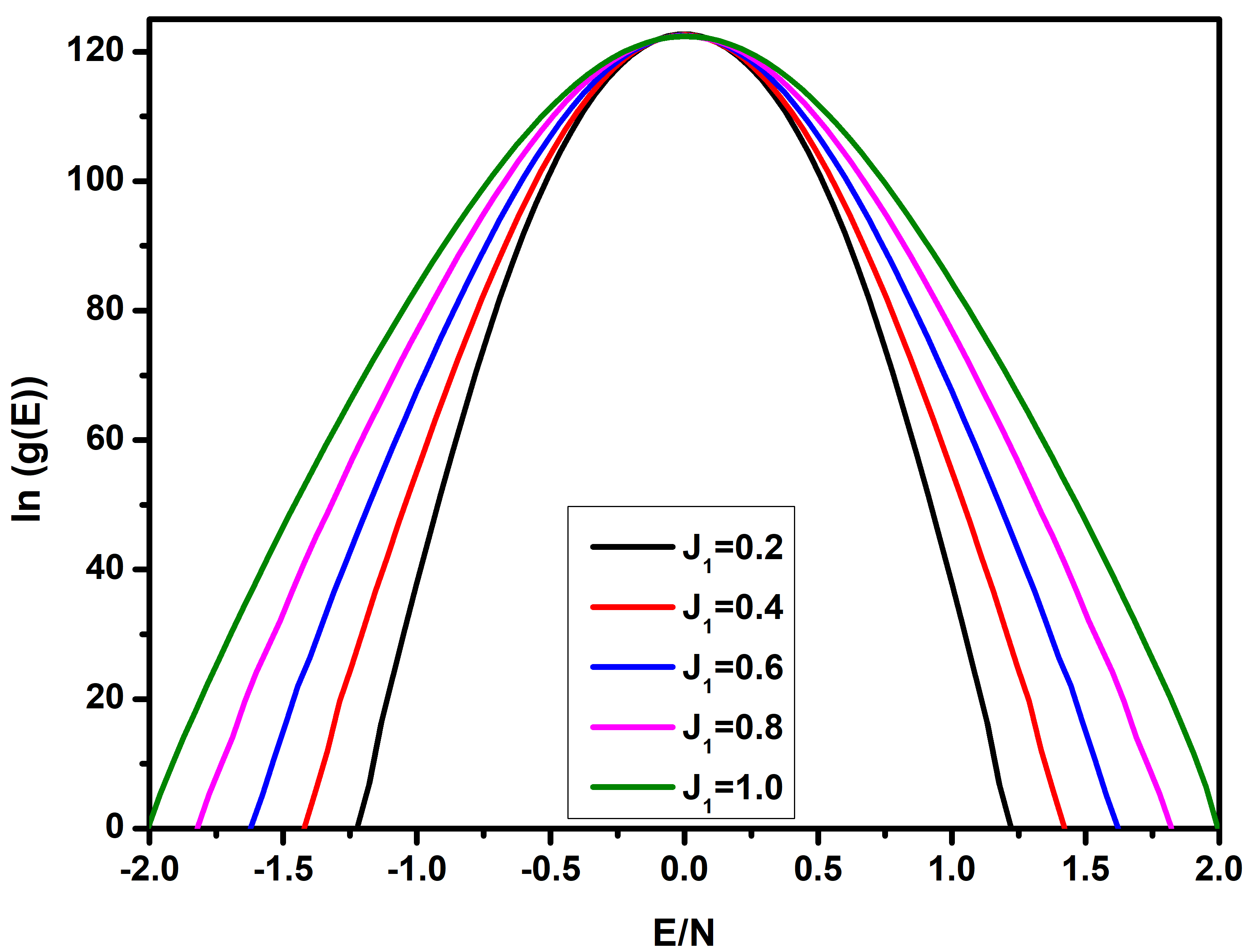}
		\caption{}
		\label{f2a}
	\end{subfigure}
	~
	\begin{subfigure}[b]{0.4\textwidth}
		\includegraphics[width=\textwidth]{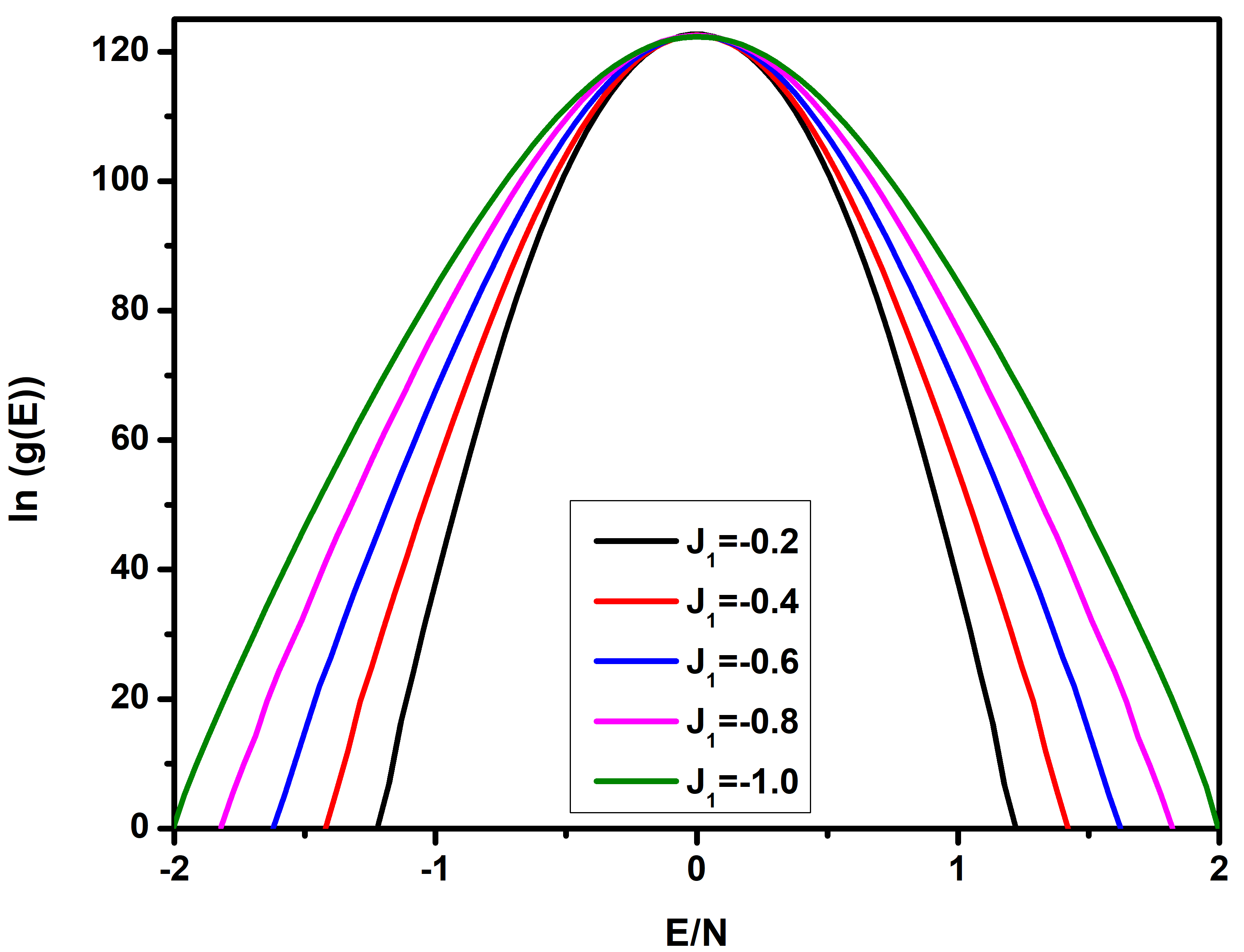} 
		\caption{}
		\label{f2b}
	\end{subfigure}
	~		
	\caption{The logarithm of DOS versus energy at $B=0$. (a) A-type AFM and (b) C-type AFM orderings.}
	\label{f2}
\end{figure}

$~~~~$In the absence of a magnetic field ($B=0$), the logarithm of DOS is computed by varying the interaction parameters. The logarithm of DOS for A-type AFM ordering is obtained by applying  a constant interaction $J_2=-1.0$ and varying the interaction $J_1$ from $0.2$ to $1.0$. Fig. \ref{f2a} represents the DOS for A-type AFM, and it is symmetric trend about zero energy. The increase in the energy span is observed by increasing the strength of the interaction parameter. The DOS of a C-type AFM is also obtained by fixing the interaction parameter $J_2$ to $1.0$ and varying $J_1$ from $-1.0$ to $-0.2$. Its DOS is also symmetric about zero energy as illustrated in Fig. \ref{f2b}. The energy span is found to decrease for C-type AFM by varying the strength of the interaction parameter. The micro-canonical entropy obtained from the estimated DOS is $S=k_B \ln g(E)$. The thermodynamic average properties are also calculated from the estimated DOS.

\begin{figure}[H]
	\centering
	\begin{subfigure}[b]{0.4\textwidth}
		\includegraphics[width=\textwidth]{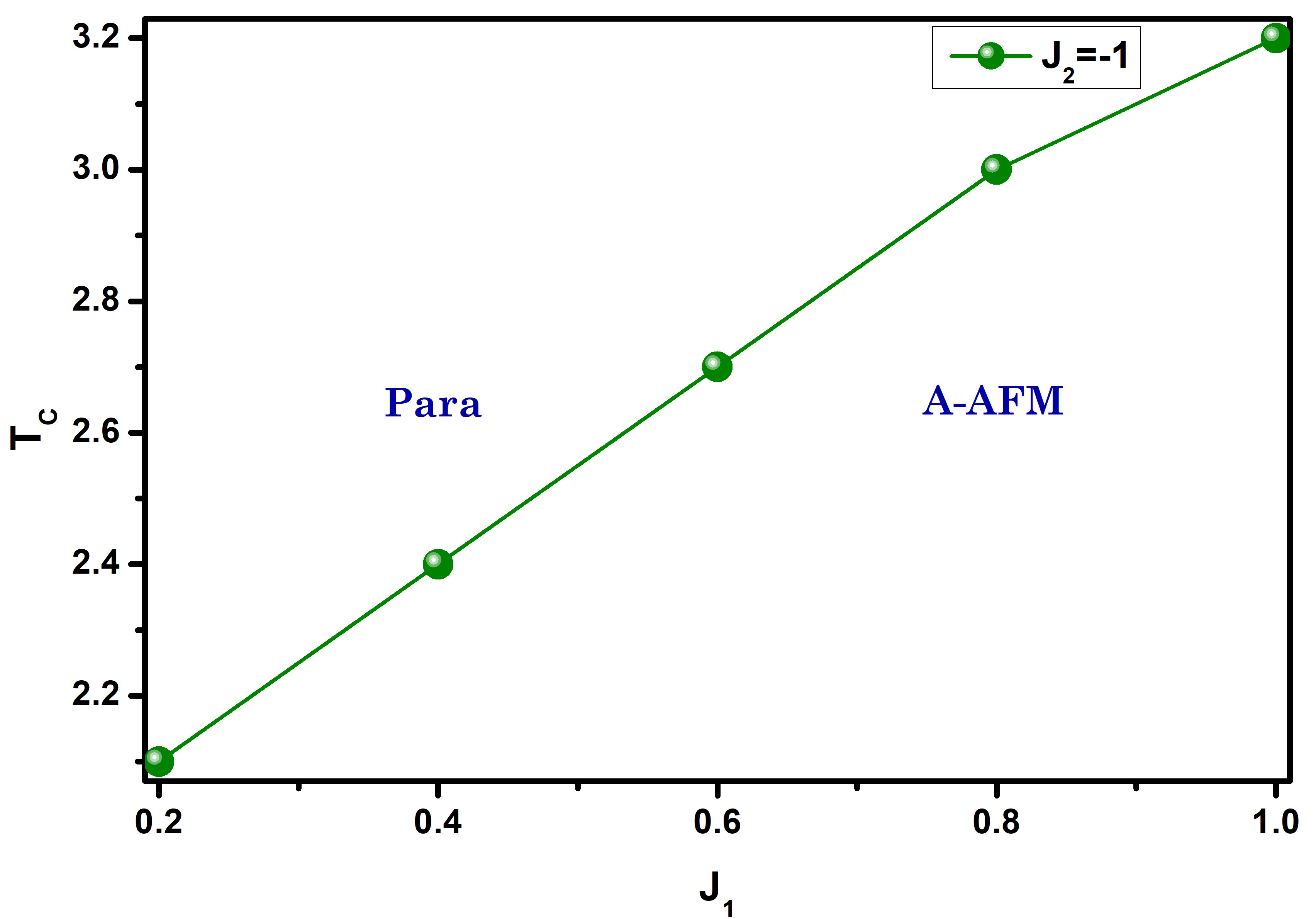}
		\caption{$ $}
		\label{f3a}
	\end{subfigure}
	~
	\begin{subfigure}[b]{0.4\textwidth}
		\includegraphics[width=\textwidth]{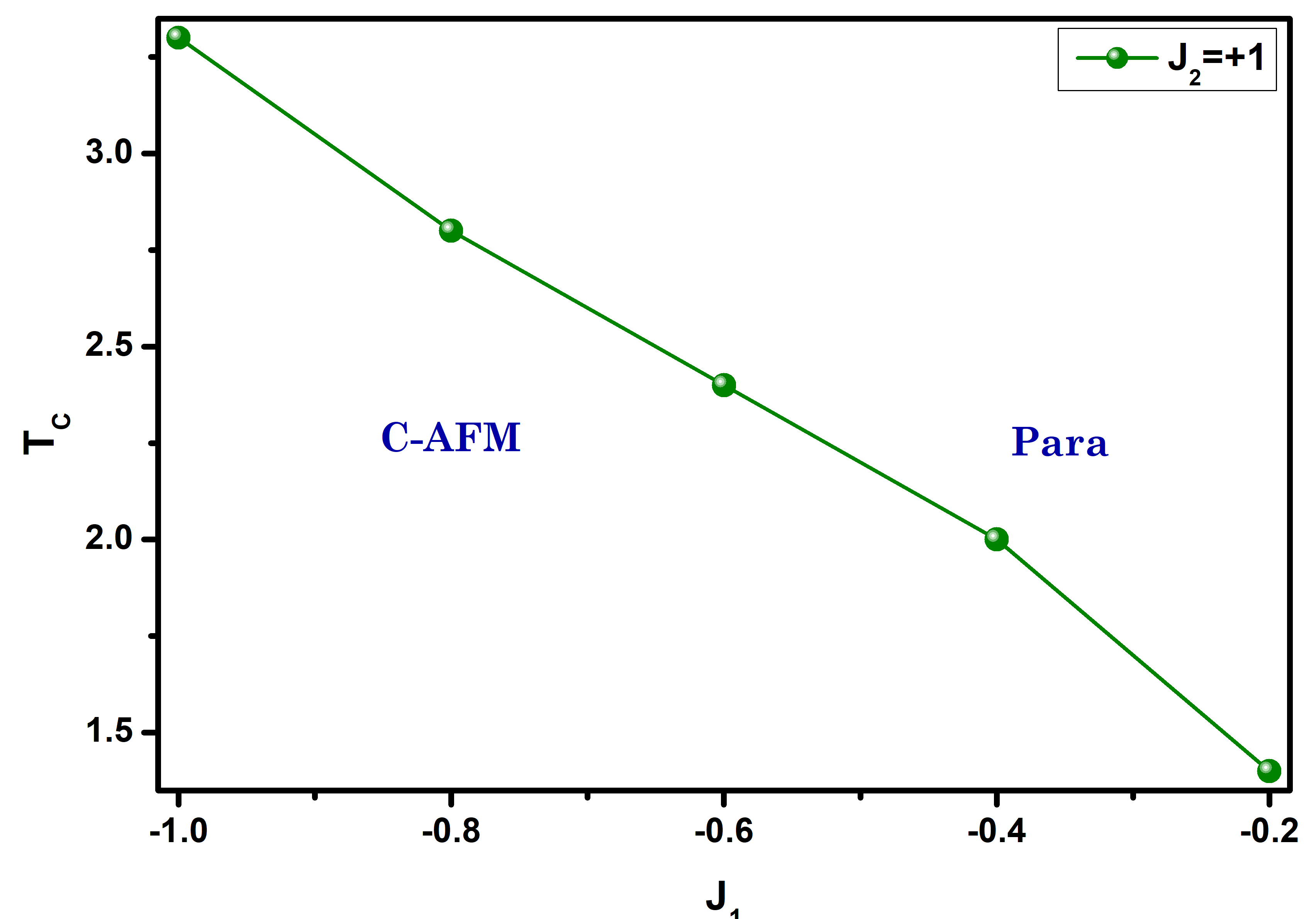}
		\caption{$ $}
		\label{f3b}
	\end{subfigure}
	~
	\caption{Phase diagram for $J_1$ versus $T_C$ at fixed $J_2$ value. (a) A-type AFM and (b) C-type AFM orderings.}
	\label{f3}
\end{figure}

The average magnetization, magnetic susceptibility, free energy, and canonical entropy are given in the Appendix. The transition temperatures ($T_C$) are collected from the susceptibility data, and the phase diagram is plotted in Fig. \ref{f3}. In Fig. \ref{f3a}, the $T_C$ increases with an increase in the interaction $J_1$. The phase curve separates the $J_1-T_C$ plane into different magnetic phases: A-type AFM phase is located under the curve, and the paramagnetic phase is found above the curve. For C-type AFM, the $T_C$ curve decreases by the increase in the magnitude of interaction $J_1$ (Fig. \ref{f3b}). Below the $T_C$ curve, the system is in C-type AFM phase, and above the $T_C$ curve, the system is in paramagnetic phase. These WL results for the opposite sign of interactions in the absence of magnetic field corroborated with our previous study using the Metropolis algorithm \cite{Elden2022Monte}.

\subsection{In the presence of magnetic field}
\begin{figure}[H]
	\centering
	\begin{subfigure}[b]{0.4\textwidth}
		\includegraphics[width=\textwidth]{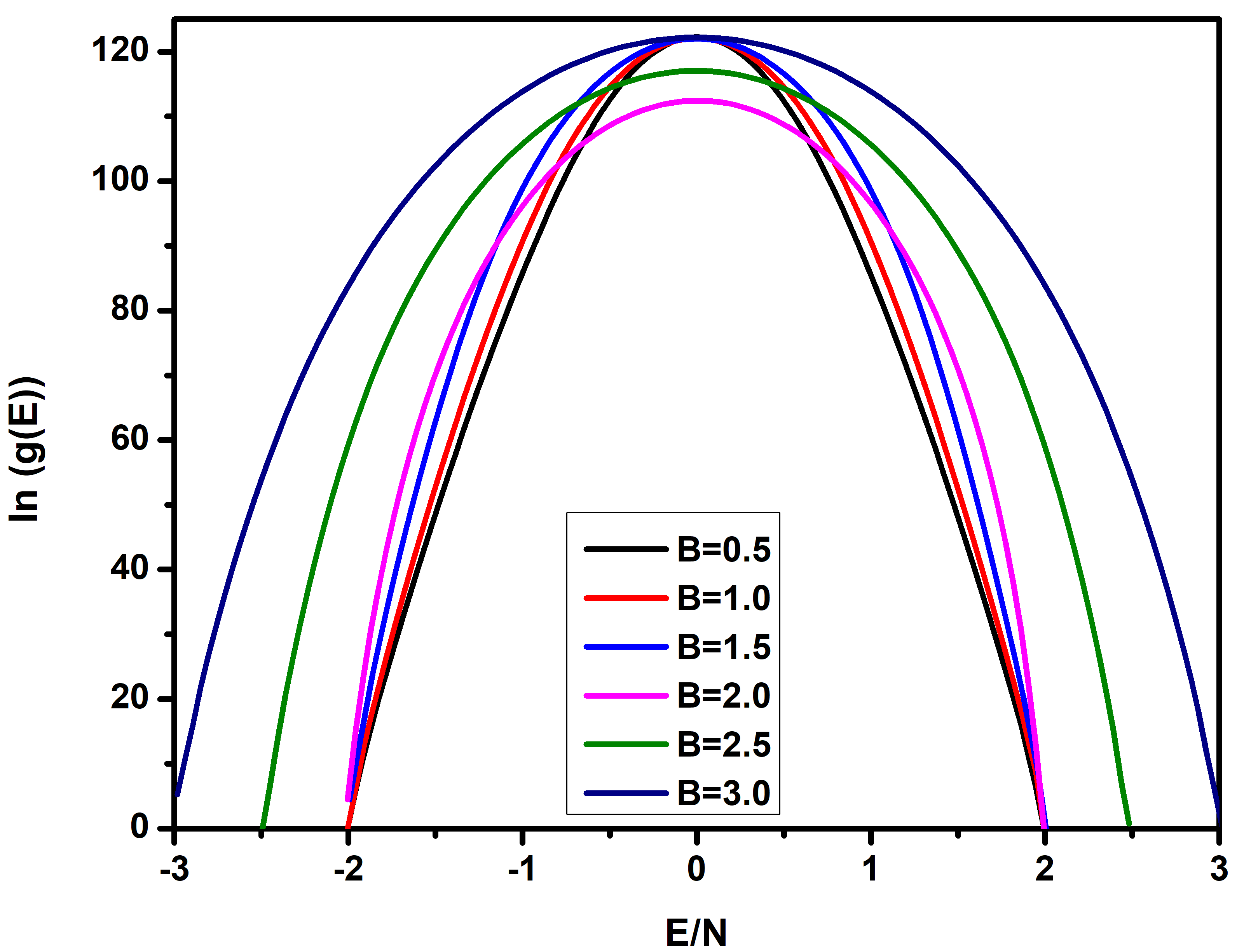}
		\caption{}
		\label{f4a}
	\end{subfigure}
	~
	\begin{subfigure}[b]{0.4\textwidth}
		\includegraphics[width=\textwidth]{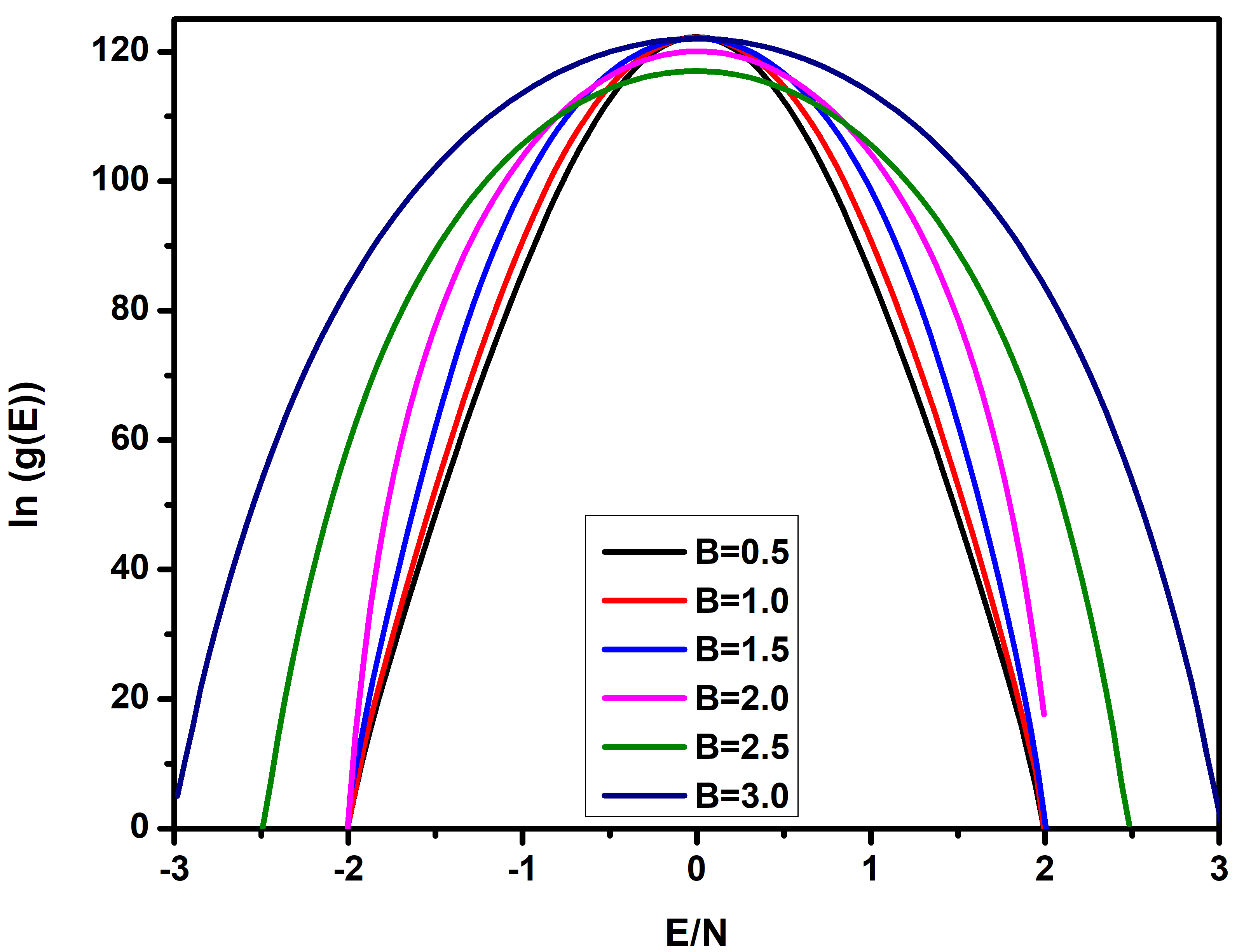} 
		\caption{}
		\label{f4b}
	\end{subfigure}
	~		
	\caption{The logarithm of DOS versus energy with varying $B$. (a) A-type AFM and (b) C-type AFM orderings.}
	\label{f4}
\end{figure}

$~~~~$In our further simulation work, the external magnetic field is applied to the system with fixed interaction strengths: for A-type AFM, $J_1=+1, J_2=-1$ and for C-type AFM, $J_1=-1, J_2=+1$. The system shows some interesting results in the presence of magnetic field. The field dependence of the logarithm of DOS is obtained for A-type AFM and C-type AFM, which is depicted in Figs. \ref{f4a} and \ref{f4b} respectively. The shape of (logarithm) DOS is also symmetrical about zero-energy with the presence of magnetic field. In our previous study, the asymmetric shape of DOS was observed for the same sign of interaction ($J_1=+1; J_2=+1$ and $J_1=-1, J_2=-1$) under the external magnetic field \cite{Elden2022Monte}. The present  result showed that the opposite sign of interactions should  preserve the symmetric trend of DOS even in the presence of applied magnetic field.  Also, the energy span remains constant at ($-2$ and $+2$) for the magnetic field 
$B \leq 2.0$. The lower energy value $E=-2$ corresponds to the ground state energy and the results suggest that the ground state energy remains same for the magnetic field $B\leq 2.0$. The energy span expands for further increase in the magnetic field ($B>2.0$). These common observations are existing in both A-type AFM and C-type AFM orderings.  The energy span increases with an increase in the applied magnetic field can be an indication of change in the phase behavior for $B >2.0$.

\begin{figure}[H]
	\centering
	\begin{subfigure}[b]{0.41\textwidth}
		\includegraphics[width=\textwidth]{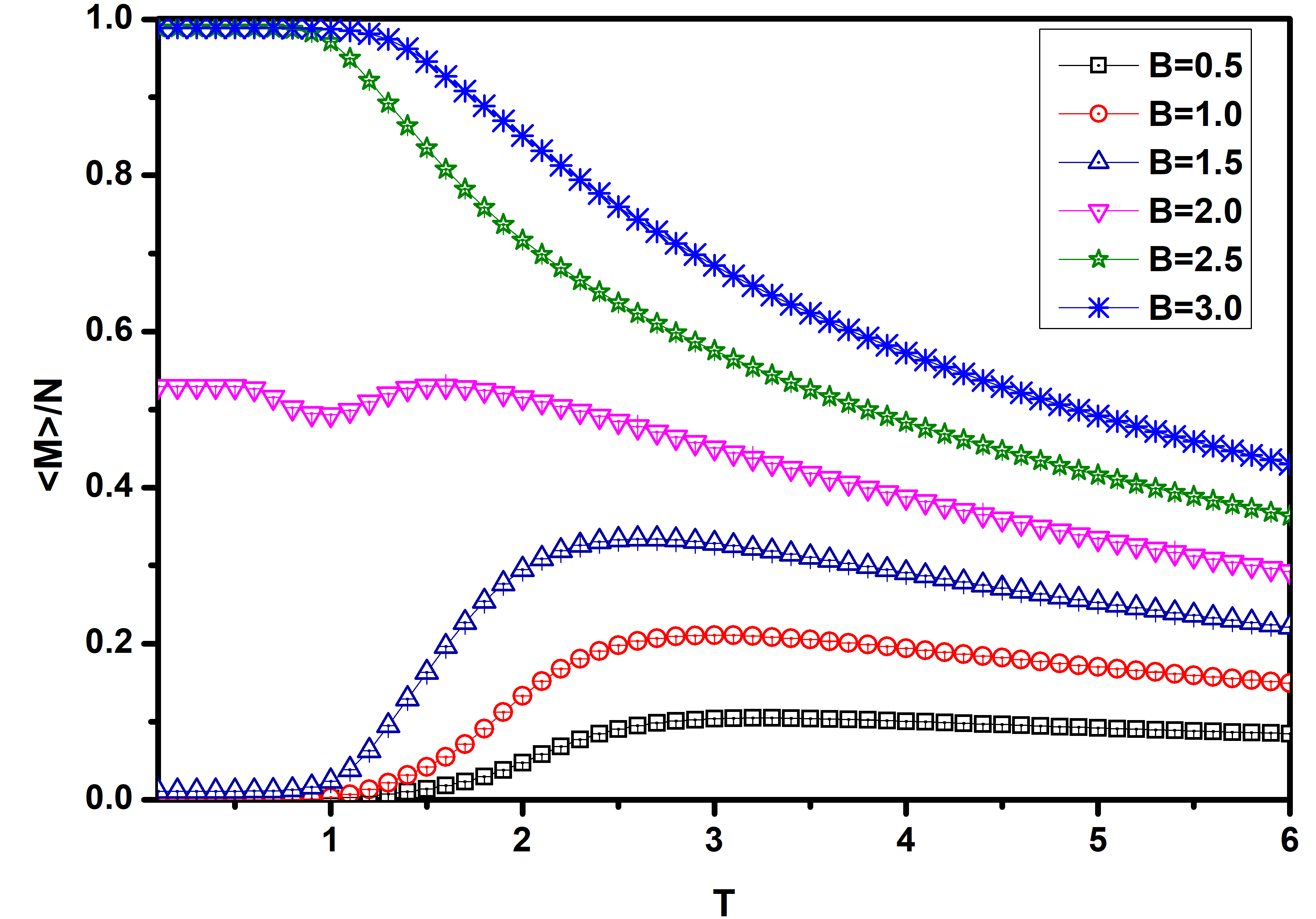}
		\caption{$ $}
		\label{f5a}
	\end{subfigure}
	~
	\begin{subfigure}[b]{0.4\textwidth}
		\includegraphics[width=\textwidth]{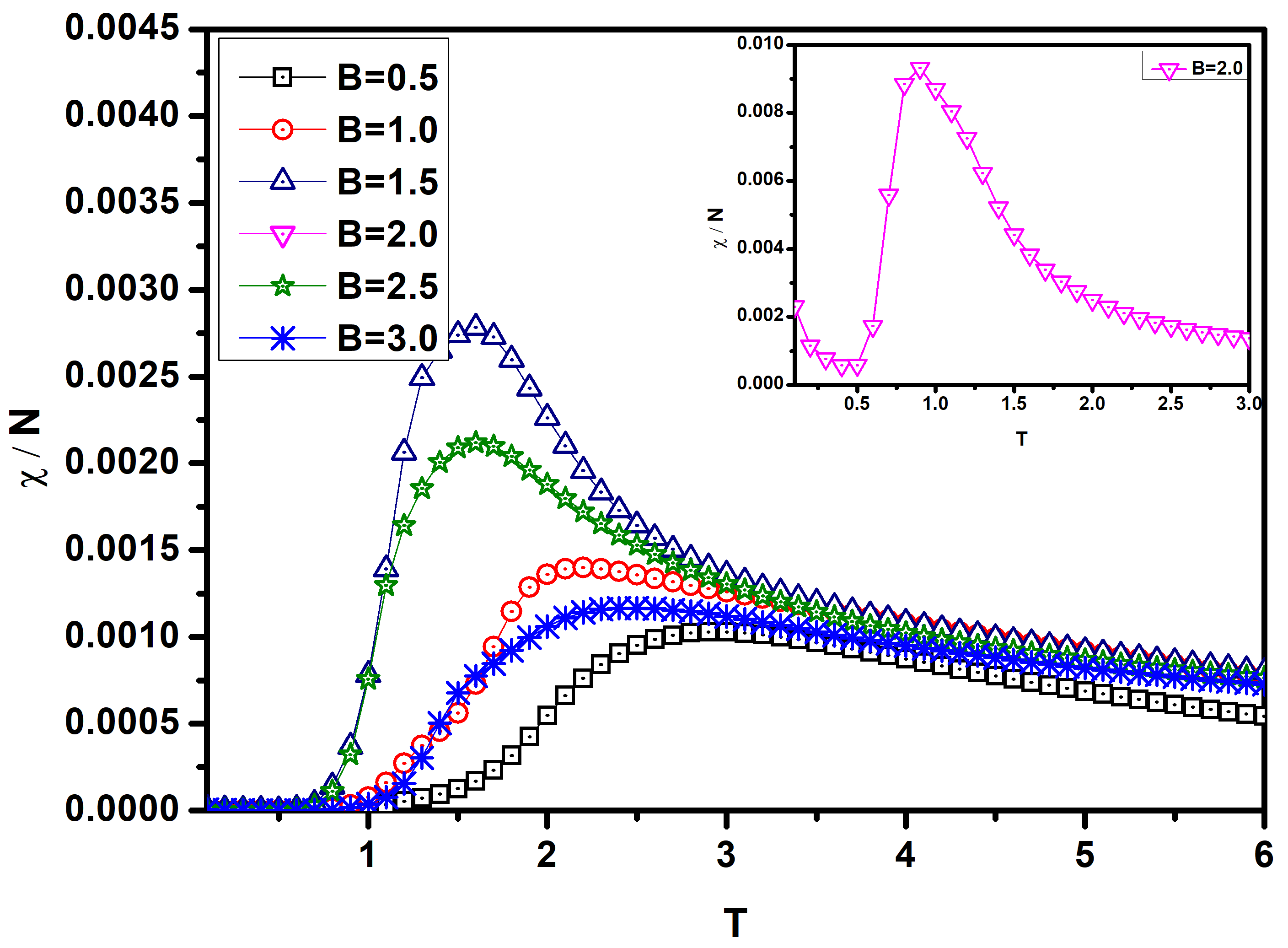}
		\caption{$ $}
		\label{f5b}
	\end{subfigure}
	~
	\caption{The field dependence of magnetic observables for A-type AFM order at various applied magnetic field strengths, (a) average magnetization and (b) susceptibility. The magnetic field varies from $0.5$ to $3.0$. The inset plot shows the susceptibility curve at $B=2.0$.}
	\label{f5}
\end{figure}

\begin{figure}[H]
	\centering
	\begin{subfigure}[b]{0.41\textwidth}
		\includegraphics[width=\textwidth]{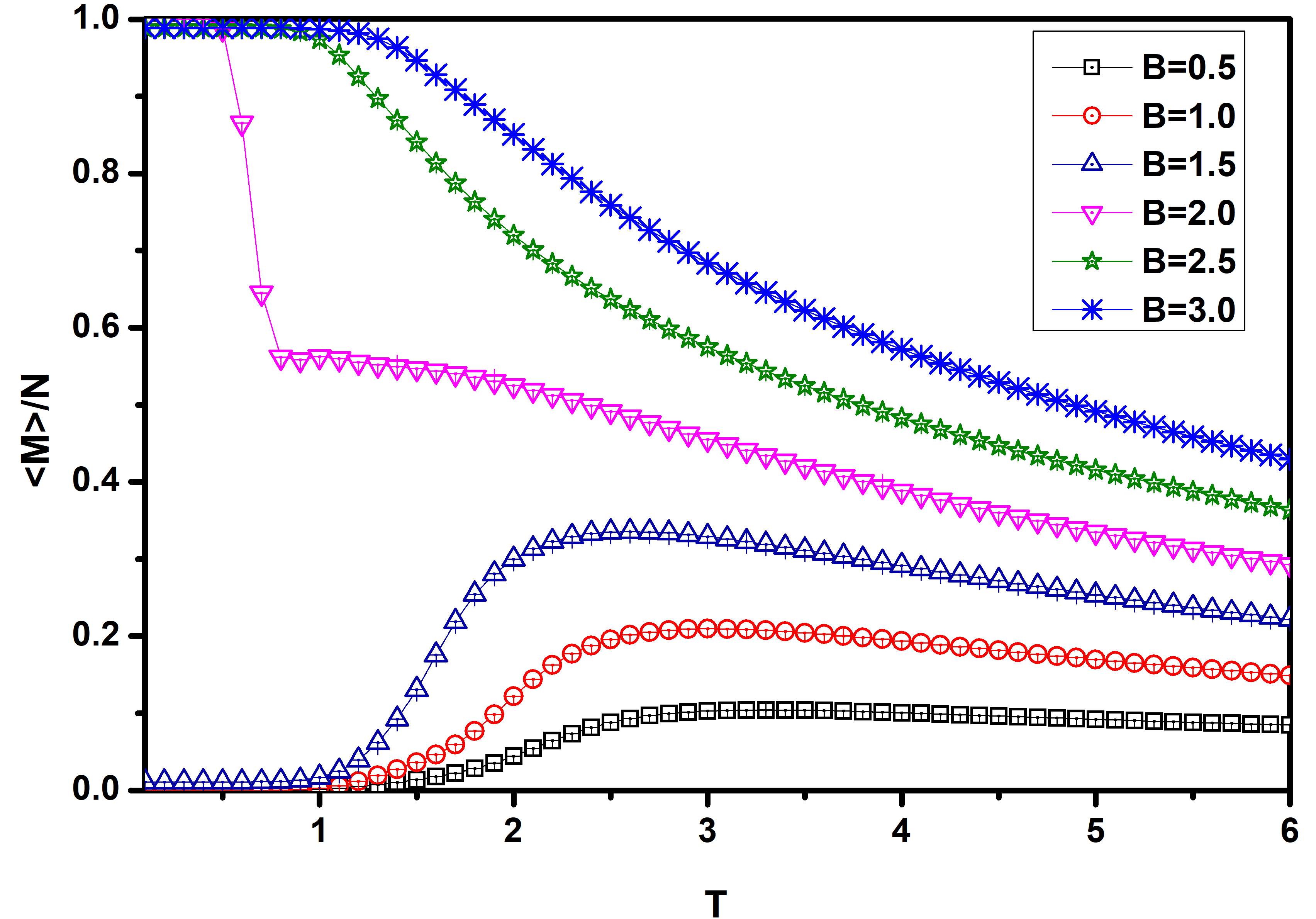}
		\caption{$ $}
		\label{f6a}
	\end{subfigure}
	~
	\begin{subfigure}[b]{0.4\textwidth}
		\includegraphics[width=\textwidth]{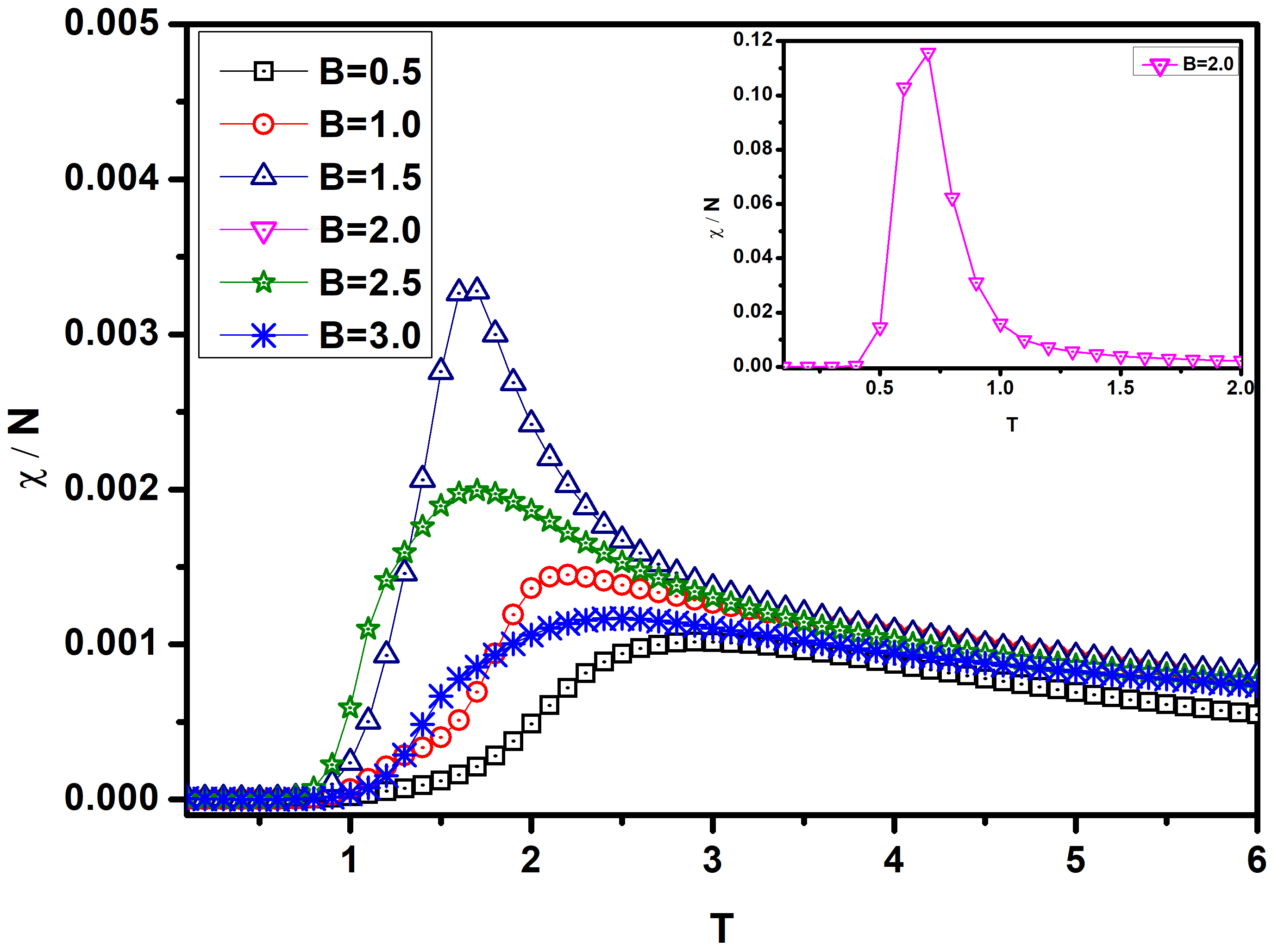}
		\caption{$ $}
		\label{f6b}
	\end{subfigure}
	~
	\caption{The field dependence of magnetic observables for C-type AFM at various applied magnetic field strengths, (a) average magnetization and (b) susceptibility. The magnetic field varies from $0.5$ to $3.0$. The inset plot shows the susceptibility curve at $B=2.0$.}
	\label{f6}
\end{figure}

In order to analyze this further, the total magnetization as a function of the magnetic field is plotted against the temperature for A-type and C-type AFM orderings, as shown in Figs. \ref{f5a} and \ref{f6a}. Both orderings exhibit zero net magnetization at low temperatures, and retain their own AFM order for magnetic field $B<2.0$. In A-type AFM, the order turns to FM ordering with an increase in the magnetic field ($B > 2.0$) even if the system is maintained at low temperatures. Fig.\ref{f5a} showed that the external magnetic field is insufficient to break the AFM order completely at low temperatures for $B \le 2.0$. Whereas, in C-type AFM, the system follows FM ordering completely when the magnetic field $B \geq 2.0 $. Fig. \ref{f6a} shows that  the total magnetization reaches to the saturated value at low temperatures as compared to Fig.\ref{f5a} 
for $B=2.0$. These results showed that, though the system is in Anti-ferromagnetic interaction, it exhibits a strong ferromagnetic ordering in the high magnetic field. The statistical error is calculated and does not exceed the size of the symbols used in the plot of entire simulated data sets. The transition temperatures are observed from the maximum value of susceptibility curves in Figs. \ref{f5b} and \ref{f6b}. The $T_C$ for both the interactions decreases until $B=2.0$, then increases for $B>2.0$. The susceptibility peak value for $B=2.0$ is higher than the rest of the $B$ values, so it is separately depicted in the inset plot of Figs. \ref{f5b} and \ref{f6b}.  In the of inset Fig. \ref{f5b} has non zero values at very low temperatures. This may be attributed to the relatively rare convergence of DOS at $B=2.0$. Hence, the system needs further careful study at $B=2.0$ in the Monte Carlo simulation. An increase in the temperature breaks the A-type/C-type AFM and FM orderings into a disordered paramagnetic ordering. These results are matched with the results in our previous work using the Metropolis algorithm \cite{Elden2022Monte}.
The obtained results are supported by the recent experimental observations of single crystals of layered materials \cite{Pakhira2022A-type} with AFM order change to paramagnetic order at high temperatures. The net magnetization has reached the saturation limit in the high magnetic field \cite{Berry2022A-type}. Our results also verified the following theoretical predictions. The $T_C$ increases for FM interaction and decreases for AFM interaction while increasing the external magnetic field \cite{Badiev2020The}. In the AFM interaction, the system’s order is changed to FM order as the external field increases at a low temperature regime \cite{Badiev2022Ground}.

\begin{figure}[H]
	\centering
	\begin{subfigure}[b]{0.4\textwidth}
		\includegraphics[width=\textwidth]{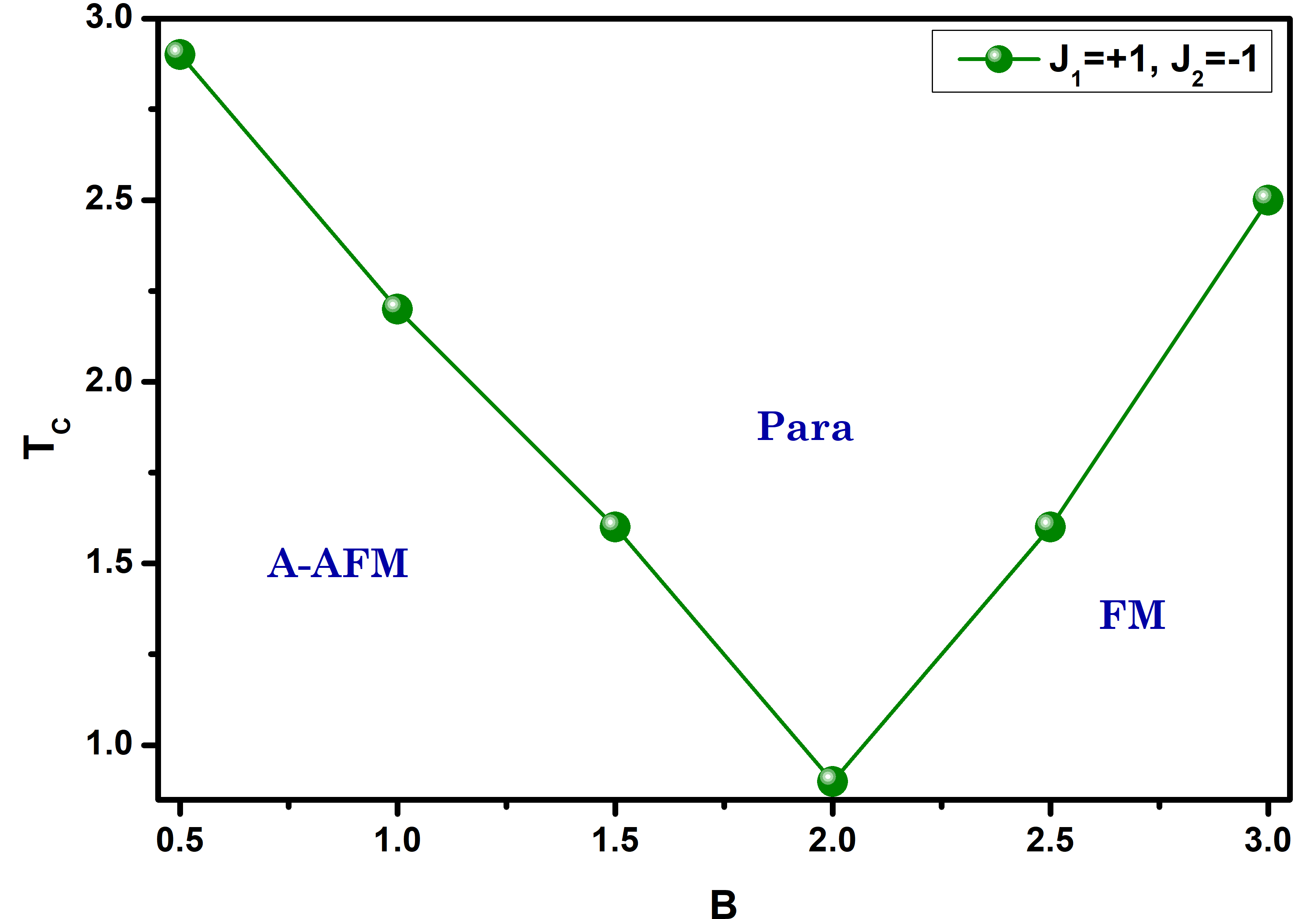}
		\caption{$ $}
		\label{f7a}
	\end{subfigure}
	~
	\begin{subfigure}[b]{0.4\textwidth}
		\includegraphics[width=\textwidth]{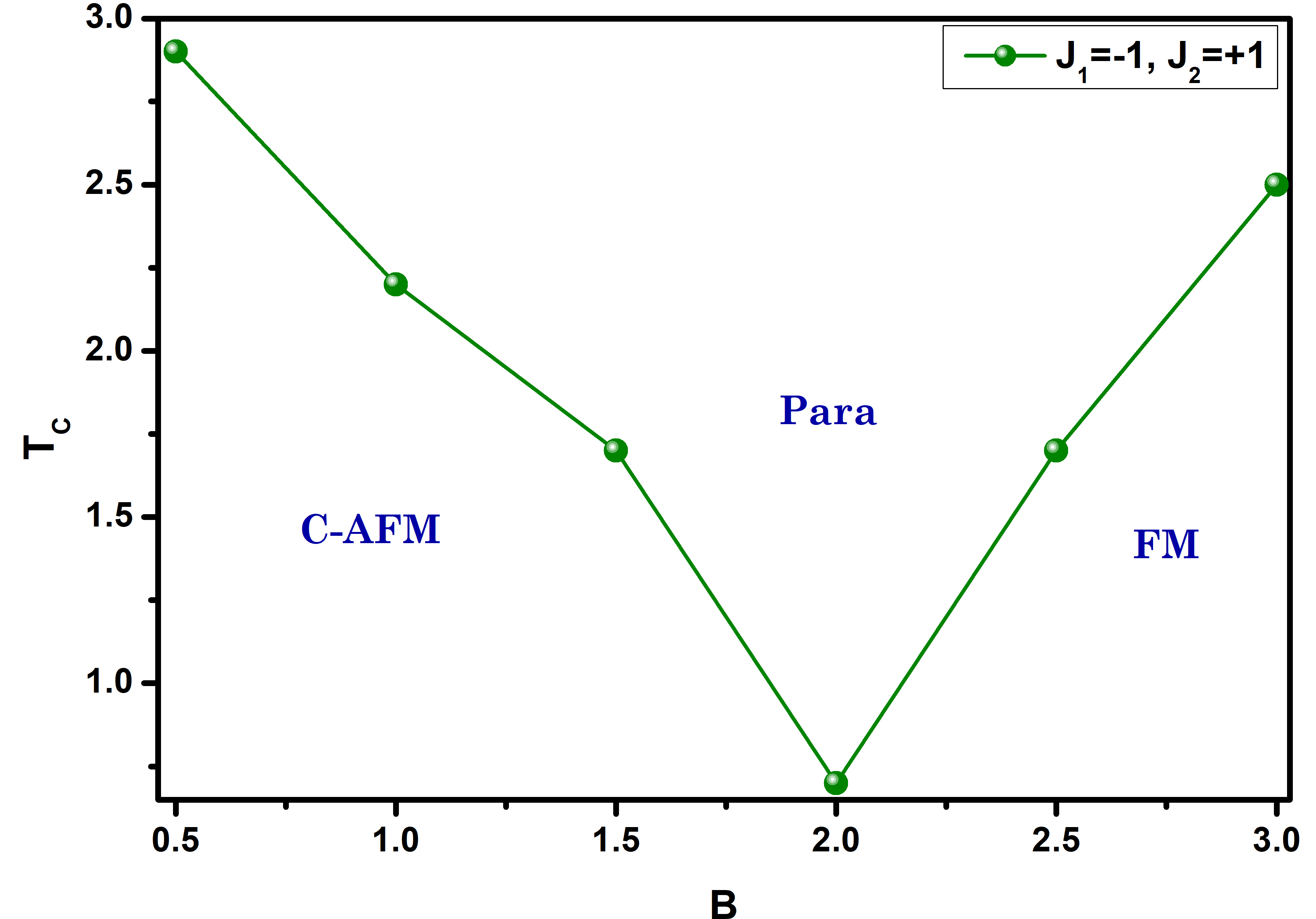}
		\caption{$ $}
		\label{f7b}
	\end{subfigure}
	~
	\caption{Phase diagram for $B$ versus $T_C$. (a) A-type AFM and (b) C-type AFM orderings.}
	\label{f7}
\end{figure}

The obtained transition temperatures are plotted in the $B-T_C$ plane as illustrated in Fig. \ref{f7}. While increasing the applied magnetic field, the $T_C$ curve gradually decreases until the field $B=2.0$, and the curve turns to increase for $B>2.0$. These findings are applicable to both A-type/C-type AFM orderings. In the case of interactions $J_1=+1, J_2=-1$, the phases of A-type AFM, FM, and paramagnetic regions are separated by the $T_C$ curve (Fig. \ref{f7a}). For the interactions $J_1=-1, J_2=+1$, $T_C$ curve divides the phases of C-type AFM, FM, and paramagnetic regions (Fig. \ref{f7b}). The line that separates the different phases meet at a single point in the vicinity of $B=2.0$, which  shows that there should be coexistence of three phases around $B=2.0$.

\begin{figure}[H]
	\centering
	\begin{subfigure}[b]{0.385\textwidth}
		\includegraphics[width=\textwidth]{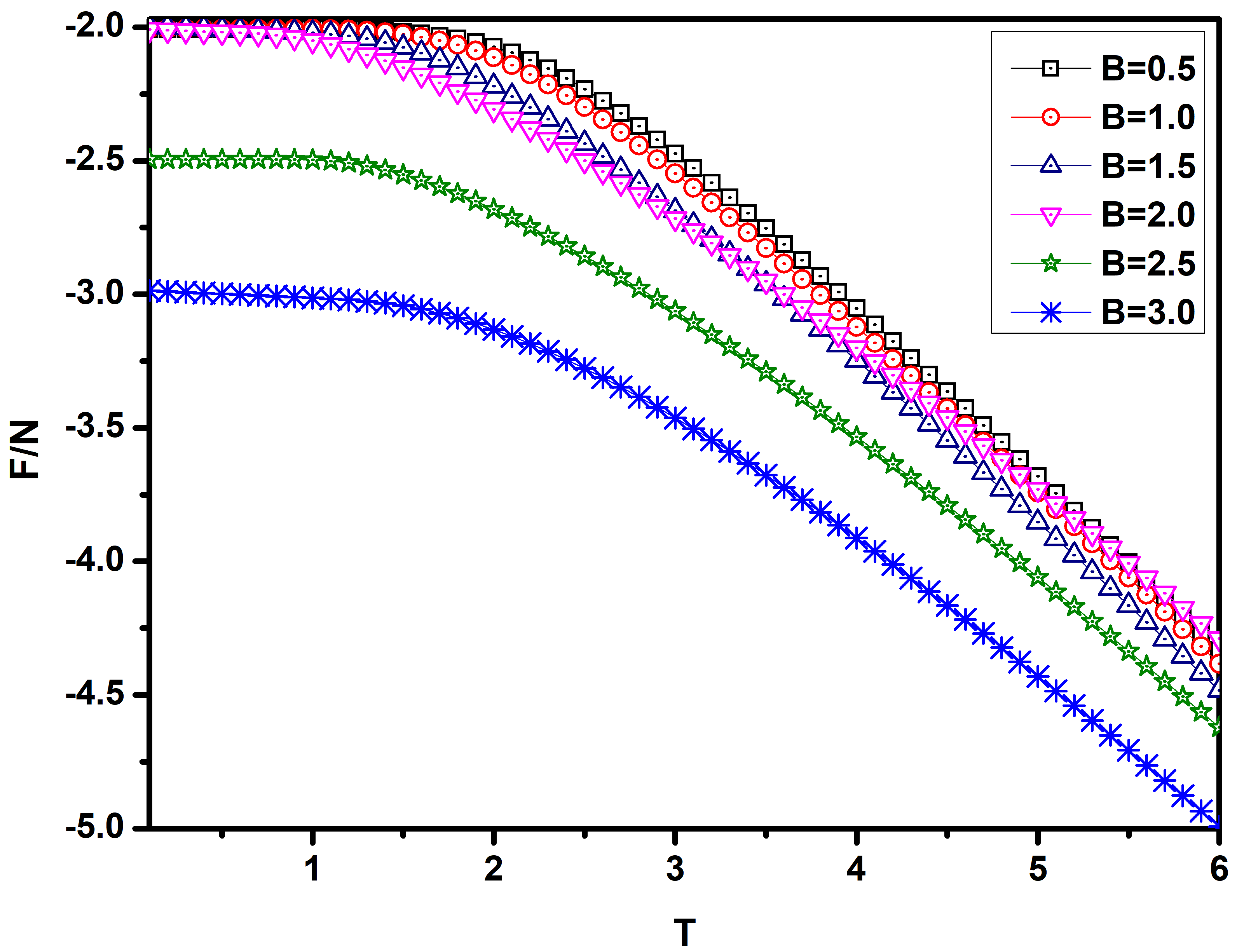}
		\caption{$ $}
		\label{f8a}
	\end{subfigure}
	~
	\begin{subfigure}[b]{0.41\textwidth}
		\includegraphics[width=\textwidth]{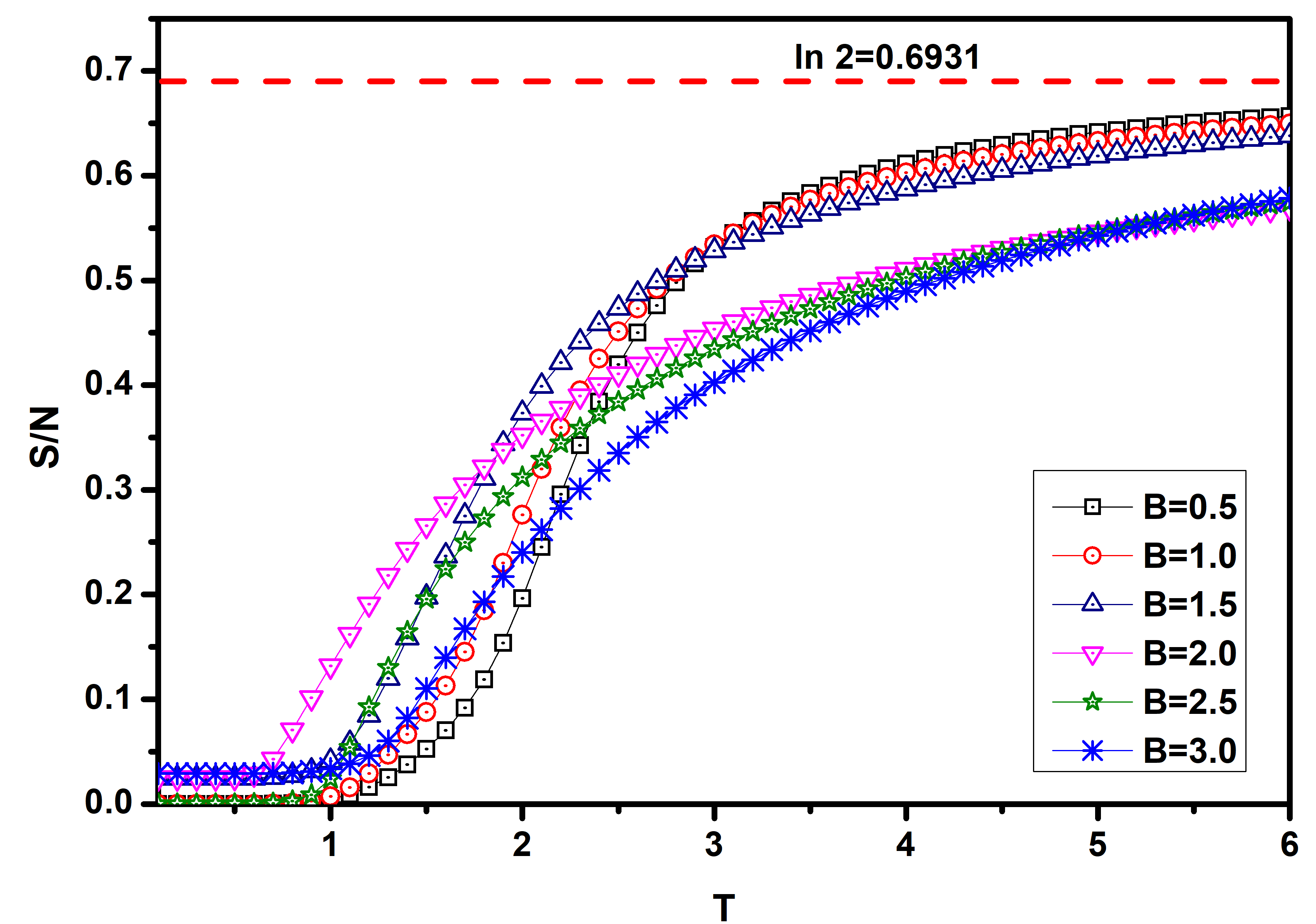}
		\caption{$ $}
		\label{f8b}
	\end{subfigure}
~
	\begin{subfigure}[b]{0.4\textwidth}
		\includegraphics[width=\textwidth]{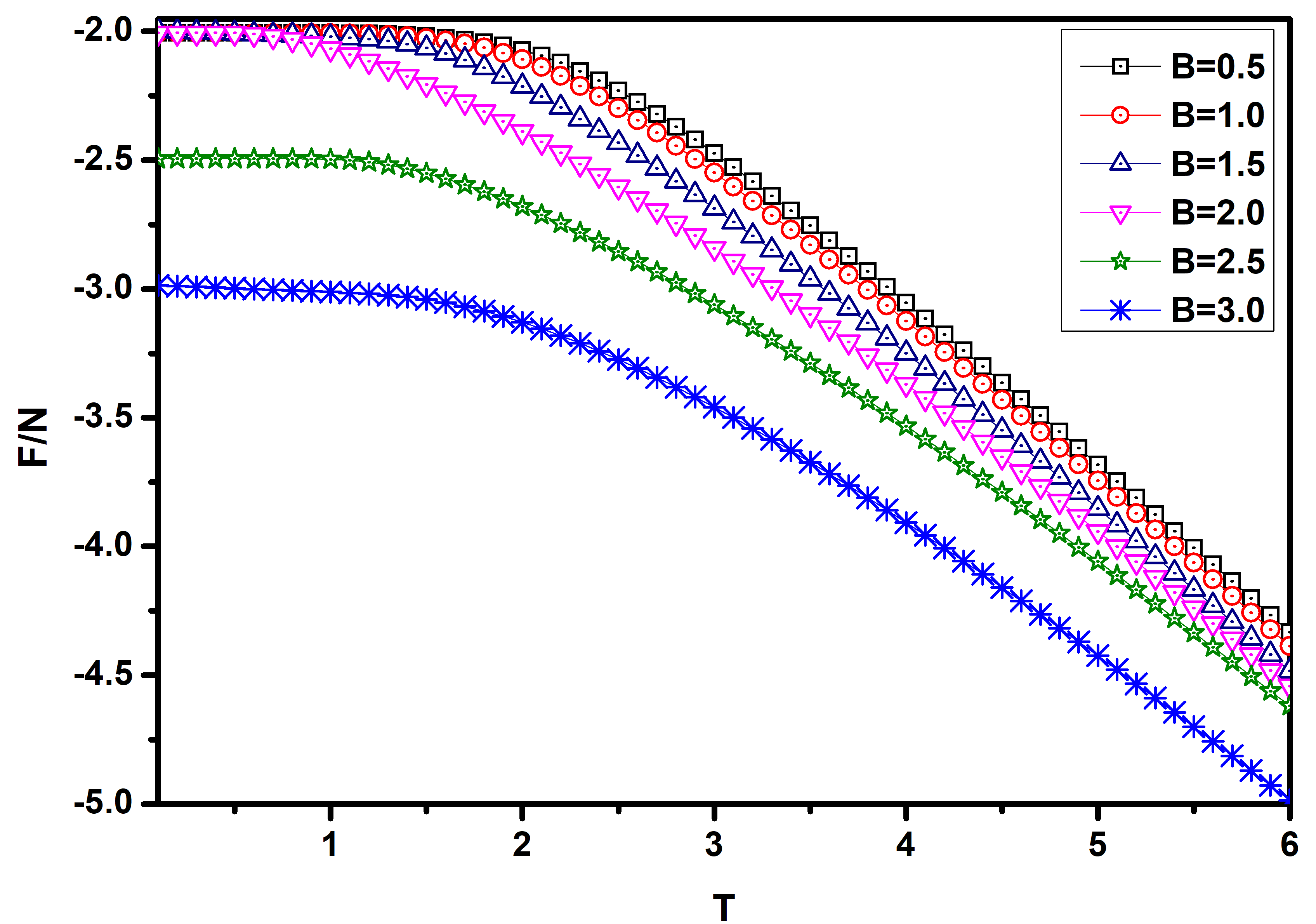}
		\caption{$ $}
		\label{f8c}
	\end{subfigure}
	~
	\begin{subfigure}[b]{0.41\textwidth}
		\includegraphics[width=\textwidth]{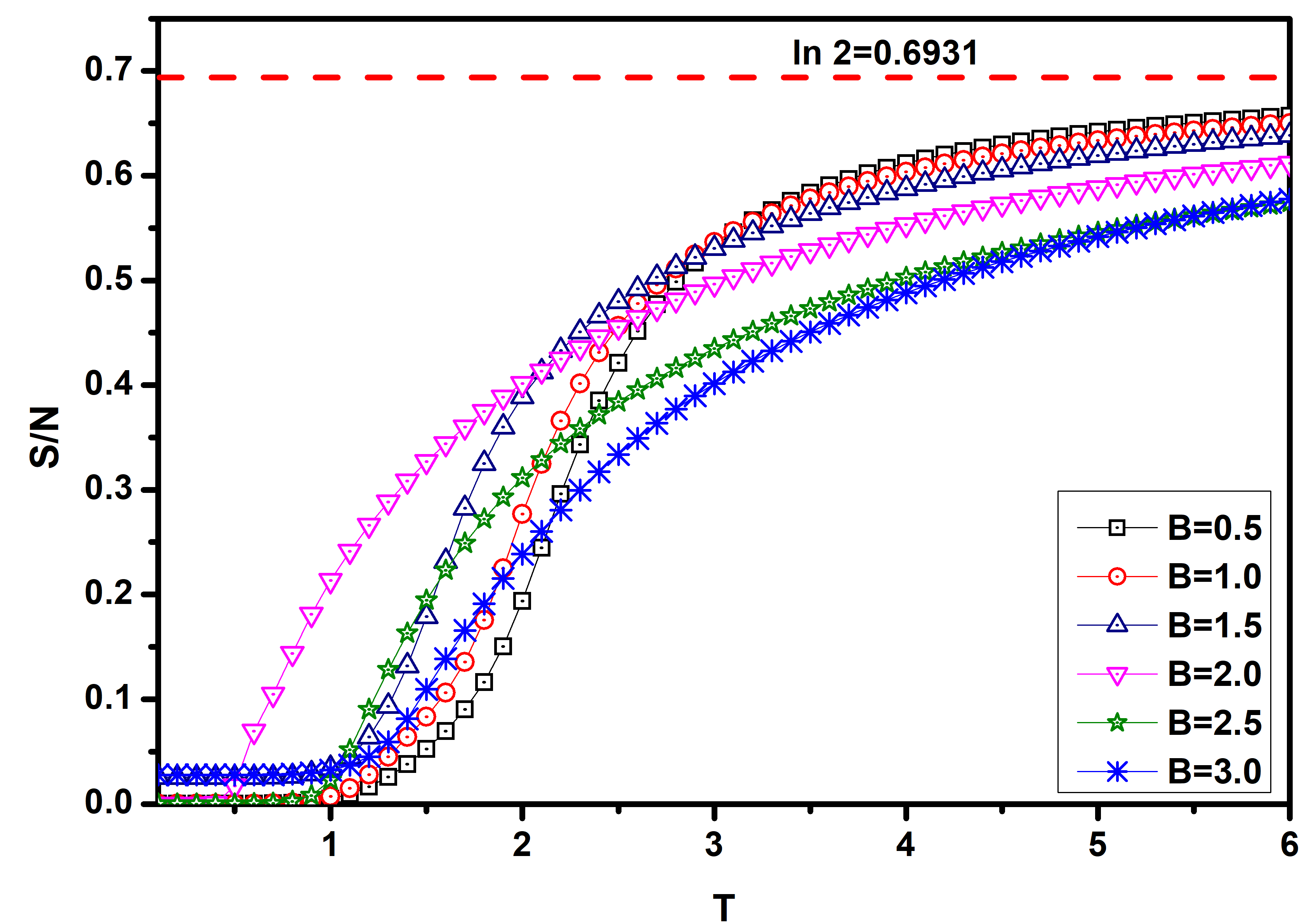}
	\caption{$ $}
		\label{f8d}
	\end{subfigure}
	\caption{Thermodynamic quantities of SWINT for various magnetic field strengths. (a) free energy and (b) entropy for A-type AFM; (c) free energy and (d) entropy for C-type AFM.}
	\label{f8}
\end{figure}

Figure \ref{f8} depicts the field dependence of free energy and entropy for the respective interactions. At low temperatures, free energy values are nearly equal to the ground state energy values observed from the density of states for corresponding interactions. Here, the value of ground state energy ($-2.0$) is constant for any anti-ferromagnetic ordering for $B<2.0$ (Figs. \ref{f8a} and \ref{f8c}).  As we discussed before, the anti-ferromagnetic nature of the system will be lost in the magnetic field $B > 2.0$, and the system will follow the FM order. So that the ground state value is increasing for the re-ordered FM state while the magnetic field $B > 2.0$.  As temperature increases, all the free energy curves gradually decreases. The observations from entropy curves clearly show that the AFM curves (for $B< 2$) and the FM curves (for $B > 2$) merge separately at high temperatures. There is no discontinuity in the entropy curve, which confirms that the system exhibits the second-order phase transition \cite{Landau2004A_New}. The theoretical magnetic entropy is calculated using the following formula, $S_m=R \ln(2S+1)$, where $R=N k_B$ is the universal gas constant, and $S$ is the spin value of the system \cite{Muthu2019GD, Midya2016Large}. The value of $S_m /N$ is calculated using $S=\frac{1}{2}$; the obtained $S_m /N=\ln 2$ is included in the entropy plot as a dotted red line. This theoretical value suggests that the maximum entropy value of the system is $\ln 2$. As the temperature increases, the calculated entropy gradually increases and approaches to $S_m$.

\begin{figure}[H]
	\centering
	\begin{subfigure}[b]{0.4\textwidth}
		\includegraphics[width=\textwidth]{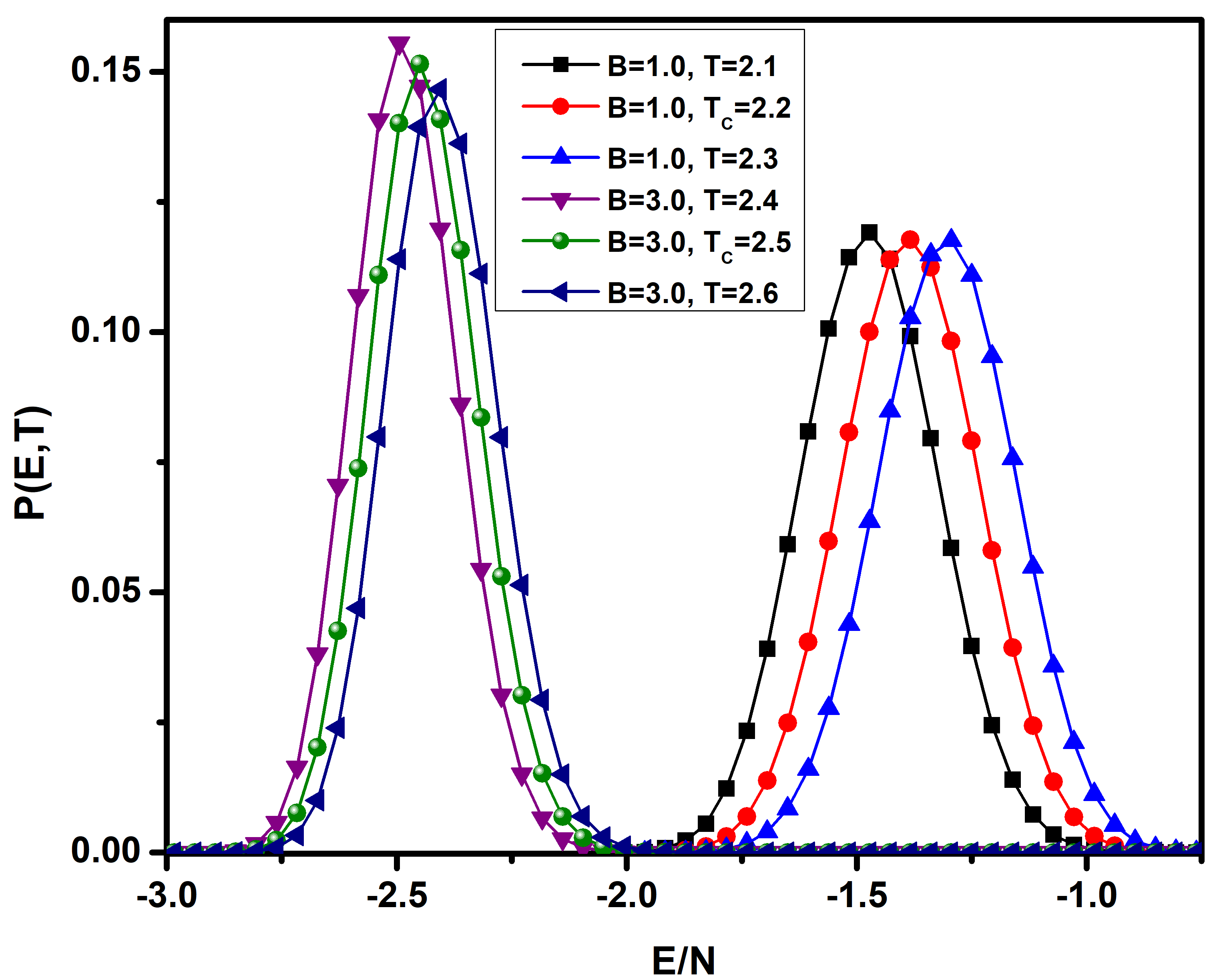}
		\caption{$ $}
		\label{f9a}
	\end{subfigure}
	~
	\begin{subfigure}[b]{0.4\textwidth}
		\includegraphics[width=\textwidth]{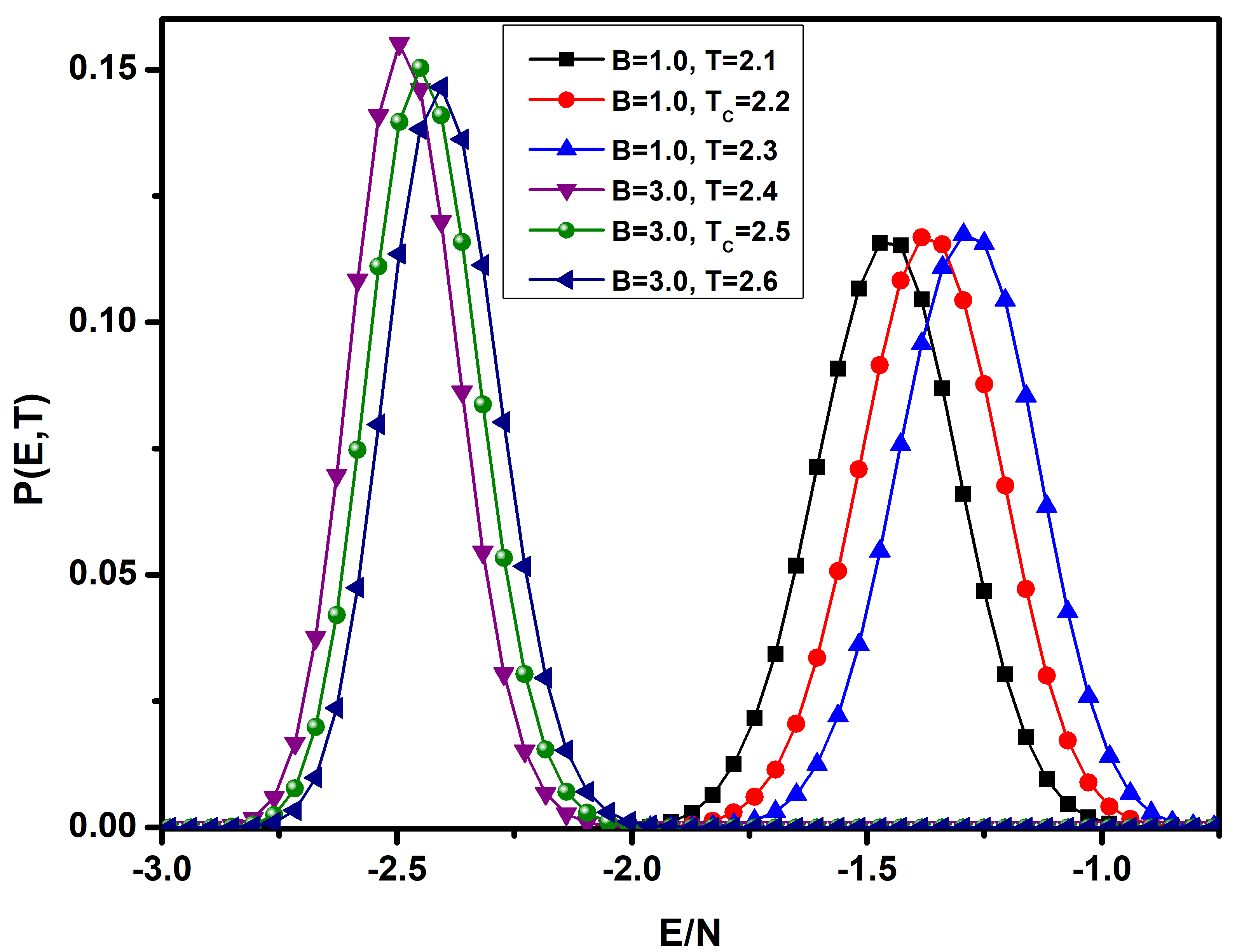}
		\caption{$ $}
		\label{f9b}
	\end{subfigure}
	\caption{The probability distribution at the vicinity of transition temperatures for two different magnetic fields. (a) A-type AFM and (b) C-type AFM orderings.}
	\label{f9}
\end{figure}

The energy distribution is also determined with the canonical probability distribution 
$P(E, T) \propto g(E)~exp[-\beta E]$ at the vicinity points of the transition temperatures in the presence of an external magnetic field \cite{Landau2004A_New}. All the configurations have been chosen based on the probability weight factor. In Fig \ref{f9a}, the distribution plot is presented for A-type AFM in the presence of magnetic fields $B=1.0$ and $B=3.0$. Furthermore, the energy distribution for C-type AFM is also given in the Fig \ref{f9b}. A single peak is observed from the distribution plots for all types of interactions and applied magnetic fields. These results again conclude that the system follows the second-order phase transition for opposite sign of interaction.

\section{Conclusion}
$~~~~$ The SWINT is investigated using the WL algorithm with opposite sign of interactions. The DOS for A-type and C-type AFM orderings are obtained in the absence and presence of an external magnetic field and its shape is observed to be always symmetrical for different magnetic fields. The external magnetic field $B$ is increased from $0.5$ to $3.0$ with fixed unit interaction strength for A-type and C-type AFM. The canonical ensemble averages ($\langle E \rangle$ and $\langle M \rangle$) for different magnetic fields are obtained from the estimated DOS. The susceptibility plots show distinct peaks that indicate the equilibrium phase transition.

The observed transition temperature increases for A-type AFM and decreases for C-type AFM with increase in $J_1$ in the absence of  magnetic field and the system's phase transits from A-type/C-type AFM (ordered phase) to paramagnetic order (disordered phase) with increasing temperature.
When the applied magnetic field increases, the AFM orderings are rearranged to FM ordering at low temperatures. All FM, A-AFM, and C-AFM order curves transit to paramagnetic order at high temperatures. The transition temperature for both types of AFM orderings decreases with increasing magnetic field until $B = 2.0$, and increases with increasing magnetic field for $B > 2.0$. The system's phase shifts from an A-type/C-type AFM to a paramagnetic phase when the external magnetic field is less than $2.0$. In the case of $B \geq 2.0$, the system switches from FM to a paramagnetic phase with increasing temperature.

The SWINT follows a second-order phase transition in the absence of a magnetic field. The second-order phase transition is also confirmed from the canonical entropy and the probability distribution
in the presence of a magnetic field.  However, the three phases  (AFM, FM and Paramagnetic) coexisting around $B=2.0$ as shown Fig.\ref{f7} indicates that there should be a tri-critical point in the vicinity of $B=2.0$. Since the spin system under study is computationally hard especially with the presence of opposite sign of interaction, the observation of the above phase coexistence with the presence of disorder and frustration \cite{ar1} must needs further careful analysis in the future.

The above analysis indicates that, by properly tuning the magnetic properties, the single-walled nanotube can be used for fabrication of new types of magnetic storage nano materials. 
Further, this study will be useful for investigating the flexomagnetoelectric effect \cite{ar2}
of low dimensional anti-ferromagnetic magnetic systems for spintronics applications in the presence of external magnetic field.

Finally, in our WL simulation, the thermodynamic observables of SWINT are calculated and analyzed  from the estimated DOS ($g(E)$) excluding the hysteresis behavior with the presence of magnetic field. In our future works, we will calculate the joint DOS ($g(E, M)$) which is the function of energy and magnetization for analyzing hysteresis behavior of the frustrated carbon nanotube structures \cite{Huang2018Anti, Ghara2022Magnet}.

\section*{Appendix: Thermodynamical average observable in the absence of magnetic field}
\begin{figure}[H]
	\centering
	\begin{subfigure}[b]{0.42\textwidth}
		\includegraphics[width=\textwidth]{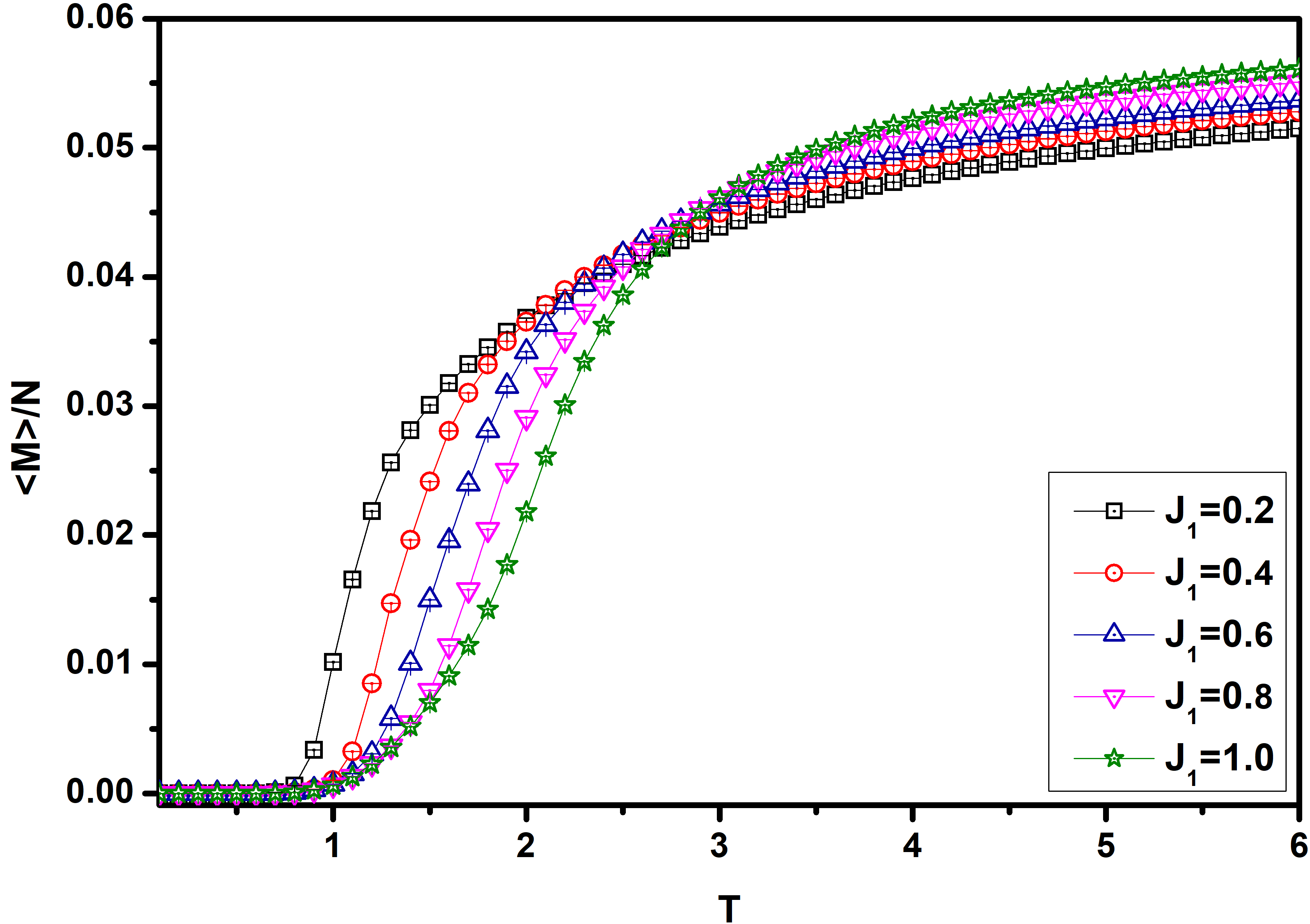}
		\caption{$ $}
		\label{f10a}
	\end{subfigure}
	~
	\begin{subfigure}[b]{0.4\textwidth}
		\includegraphics[width=\textwidth]{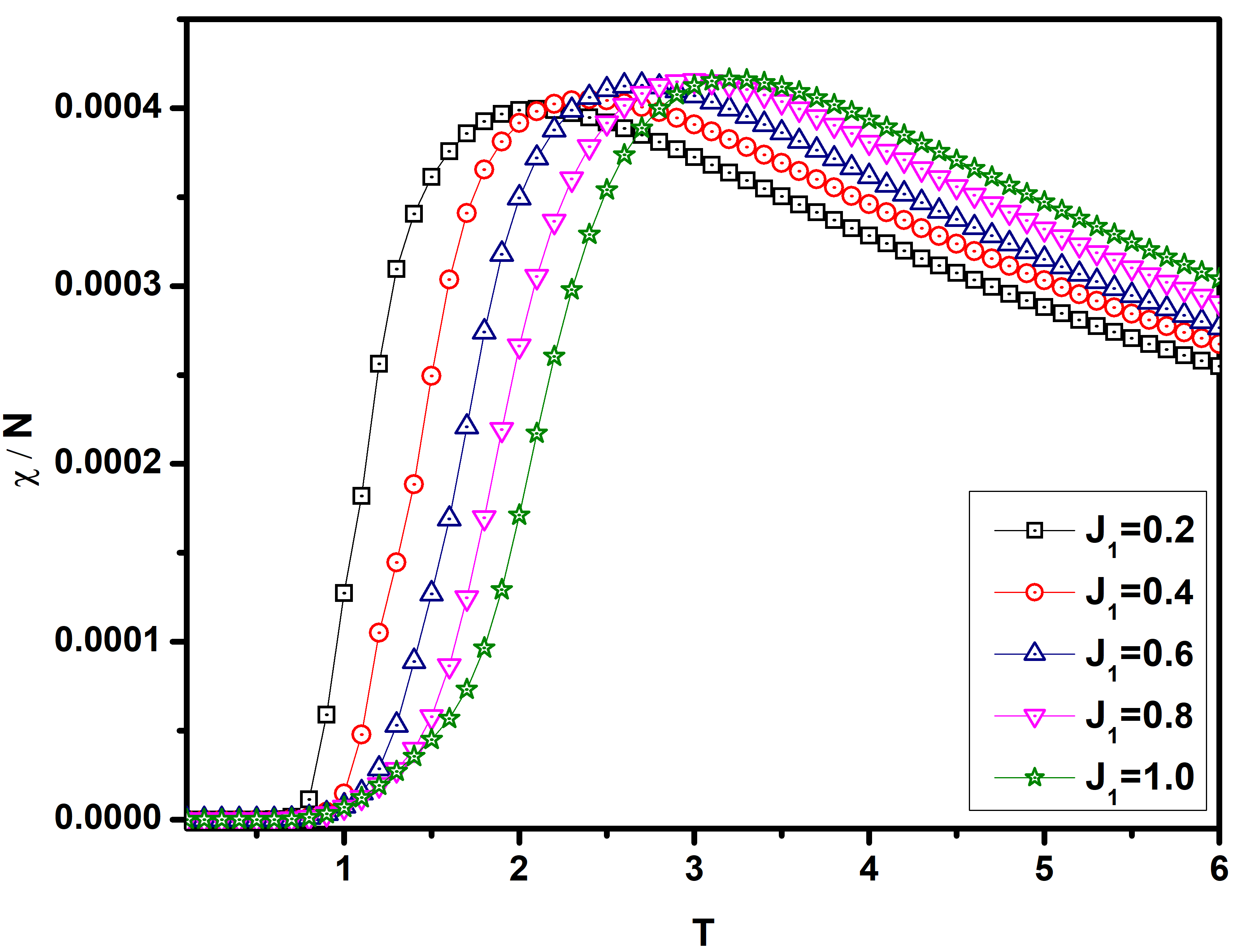}
		\caption{$ $}
		\label{f10b}
	\end{subfigure}
	~
	\caption{Magnetic observables for A-type AFM at $B=0$. (a) average magnetization and (b) susceptibility for the interactions, $0.2 \leq J_1 \leq 1.0$ and $J_2=-1.0$.}
	\label{f10}
\end{figure}

\begin{figure}[H]
	\centering
	\begin{subfigure}[b]{0.4\textwidth}
		\includegraphics[width=\textwidth]{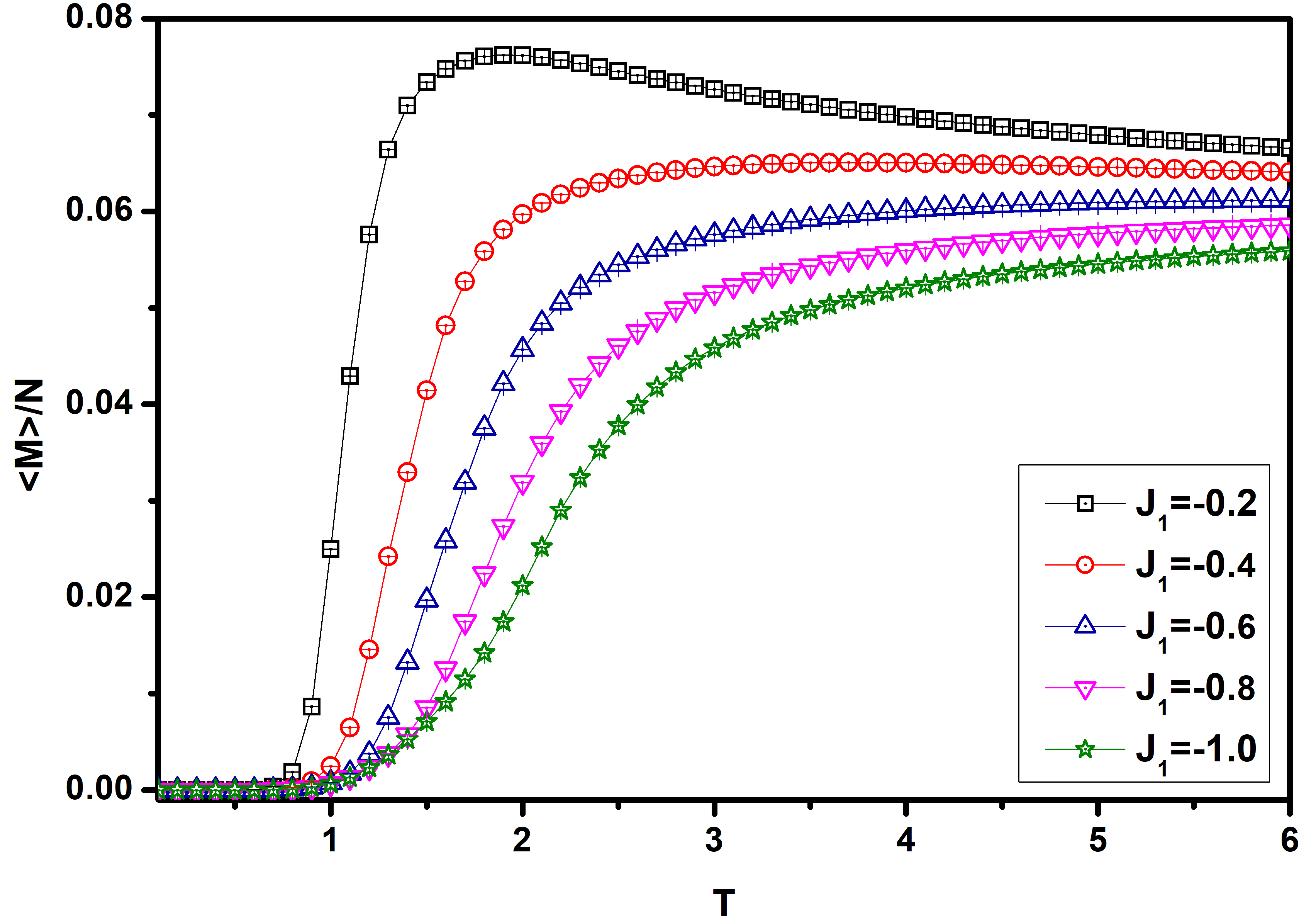}
		\caption{$ $}
		\label{f11a}
	\end{subfigure}
	~
	\begin{subfigure}[b]{0.4\textwidth}
		\includegraphics[width=\textwidth]{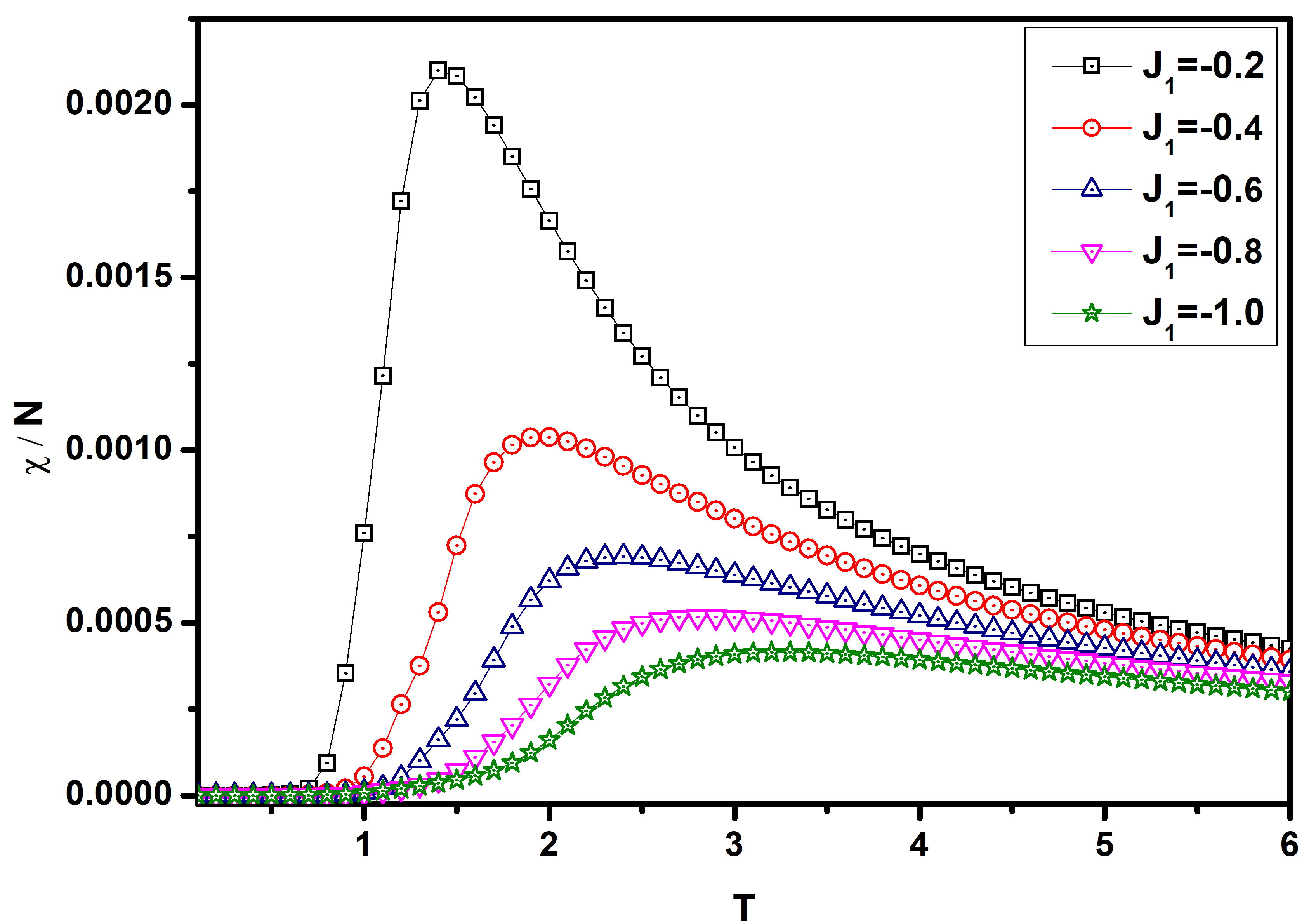}
		\caption{$ $}
		\label{f11b}
	\end{subfigure}
	~
	\caption{Magnetic observables for C-type AFM at $B=0$, (a) average magnetization and (b) susceptibility for the interactions, $-1.0 \leq J_1 \leq -0.2$ and $J_2=1.0$.}
	\label{f11}
\end{figure}

$~~~~$In the absence of magnetic field, the net magnetization is a zero at low temperatures for all anti-ferromagnetic orderings as all the spins are aligned in anti-parallel direction. The total magnetization of varying interaction strengths is depicted in Figs. \ref{f10a} and \ref{f11a} for A-type and C-type AFM orderings, respectively. The increase in temperature can disrupt the anti-parallel alignment into random directions to produce non-zero magnetization values. Thus, the total magnetization gradually increases for all interactions and reaches its saturation value. Therefore, the anti-ferromagnetic order breaks with temperature and turns to the paramagnetic order. The magnetic susceptibility curves of various interaction strengths are shown in Figs. \ref{f10b} and \ref{f11b} for A-type and C-type AFM orderings. The transition temperature are calculated from the peak in susceptibility plot. The $T_C$ value increases with increase in interaction $J_1$ from $0.2$ to $1.0$ for A-type AFM ordering (Fig. \ref{f10b}).  For C-type ordering, the $T_C$ decreases while increasing the value of interaction $J_1$ from $-1.0$ to $-0.2$ (Fig. \ref{f11b}).

\begin{figure}[H]
	\centering
	\begin{subfigure}[b]{0.4\textwidth}
		\includegraphics[width=\textwidth]{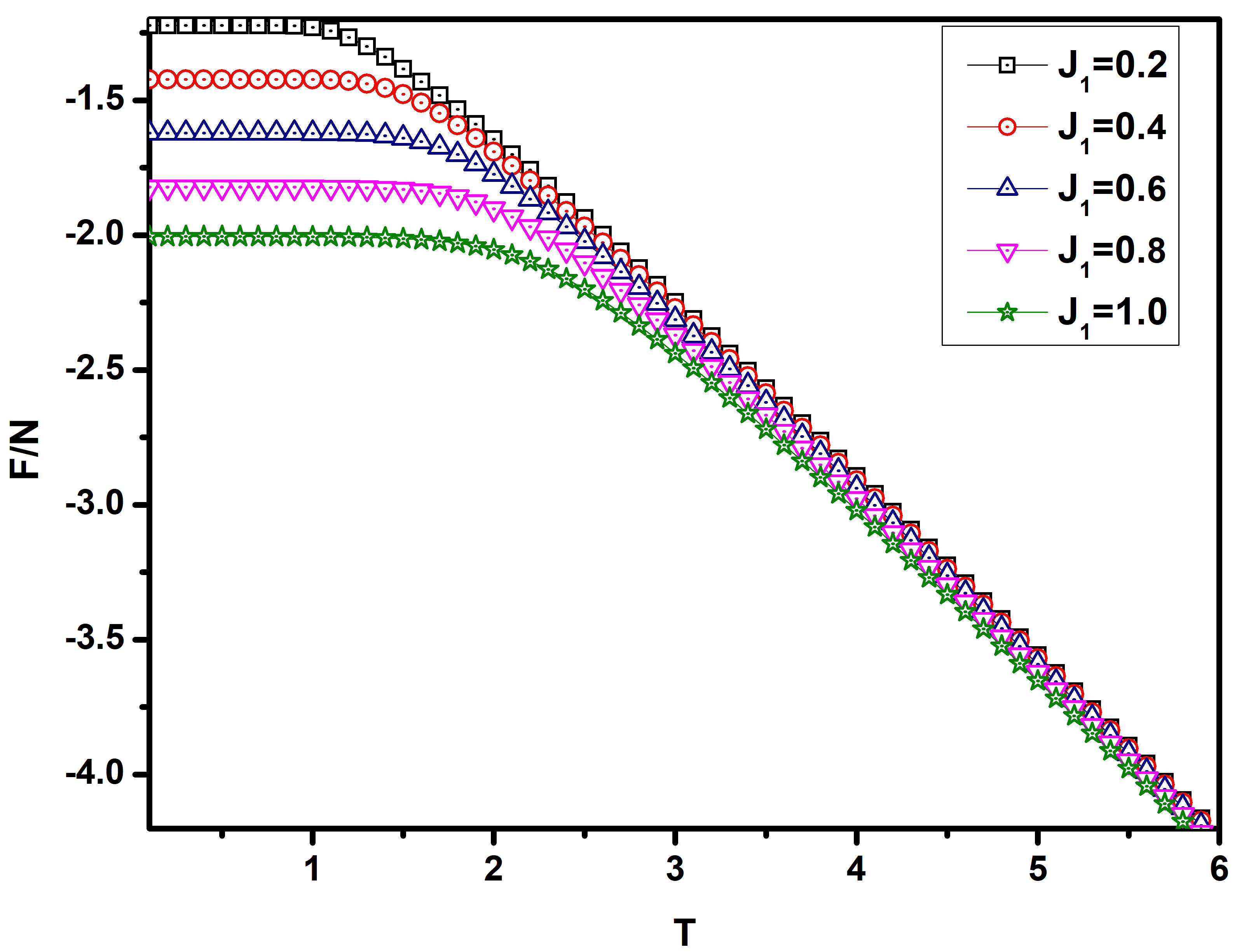}
		\caption{$ $}
		\label{f12a}
	\end{subfigure}
	~
	\begin{subfigure}[b]{0.4\textwidth}
		\includegraphics[width=\textwidth]{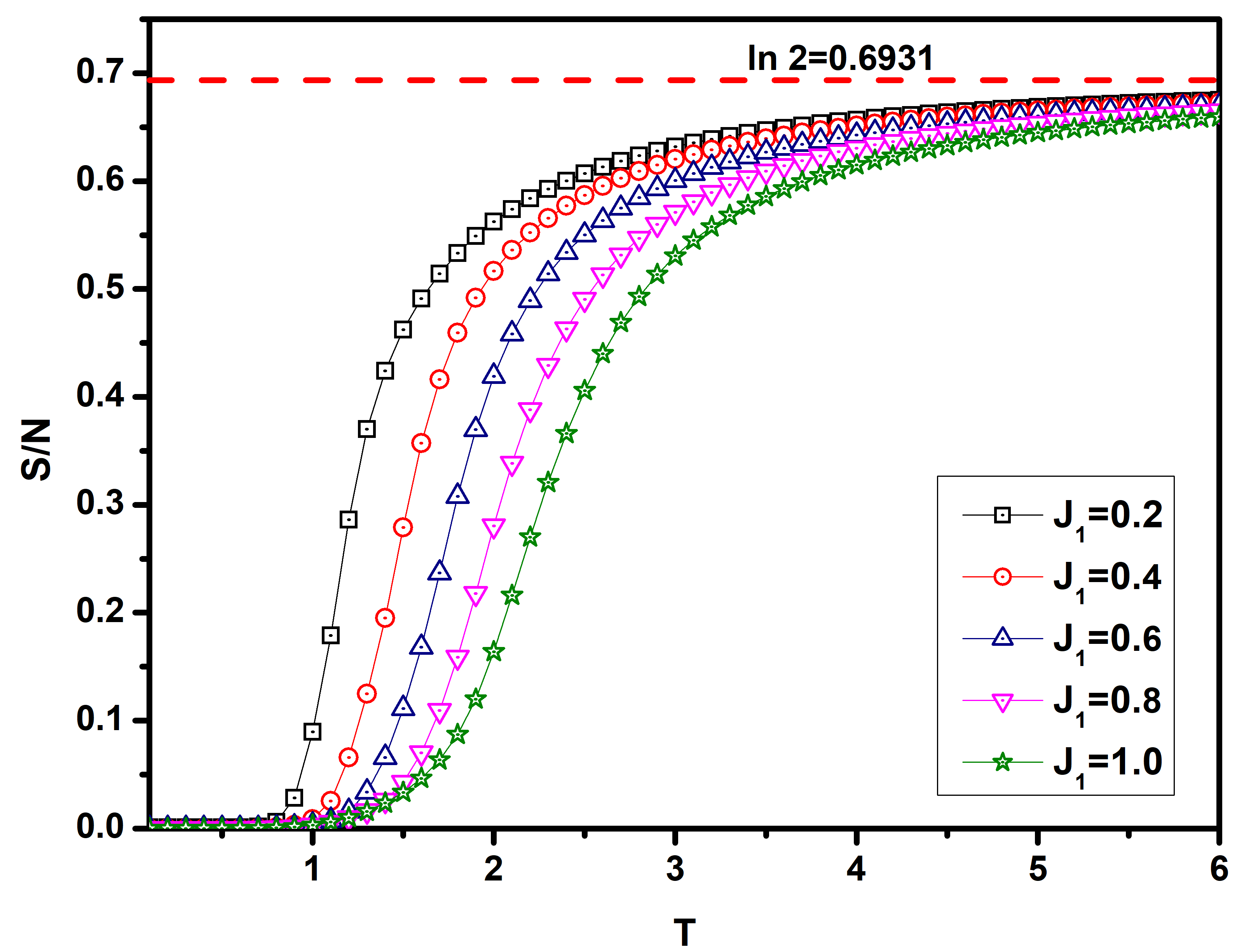}
		\caption{$ $}
		\label{f12b}
	\end{subfigure}
~
	\begin{subfigure}[b]{0.4\textwidth}
		\includegraphics[width=\textwidth]{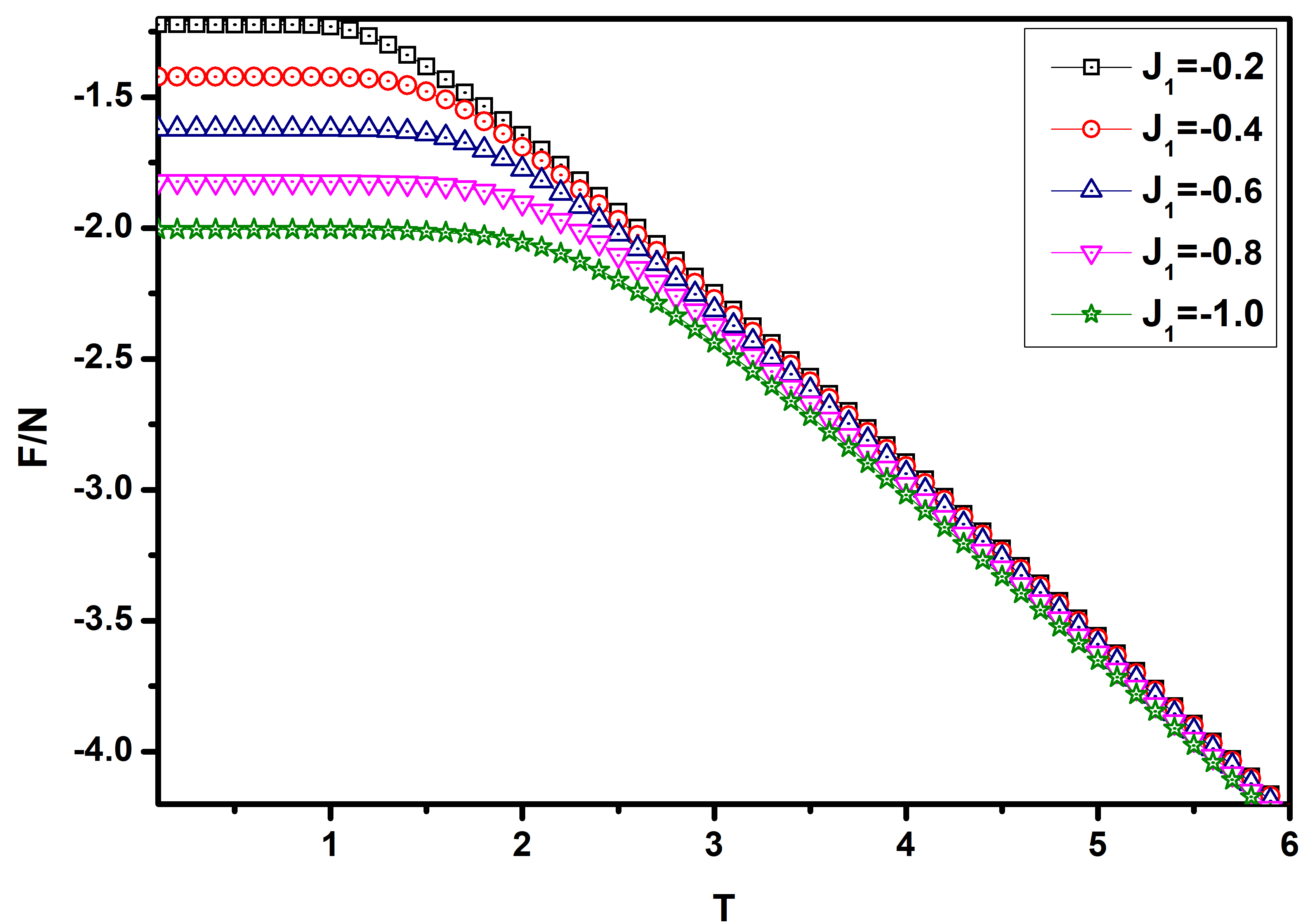}
		\caption{$ $}
		\label{f12c}
	\end{subfigure}
	~
	\begin{subfigure}[b]{0.4\textwidth}
		\includegraphics[width=\textwidth]{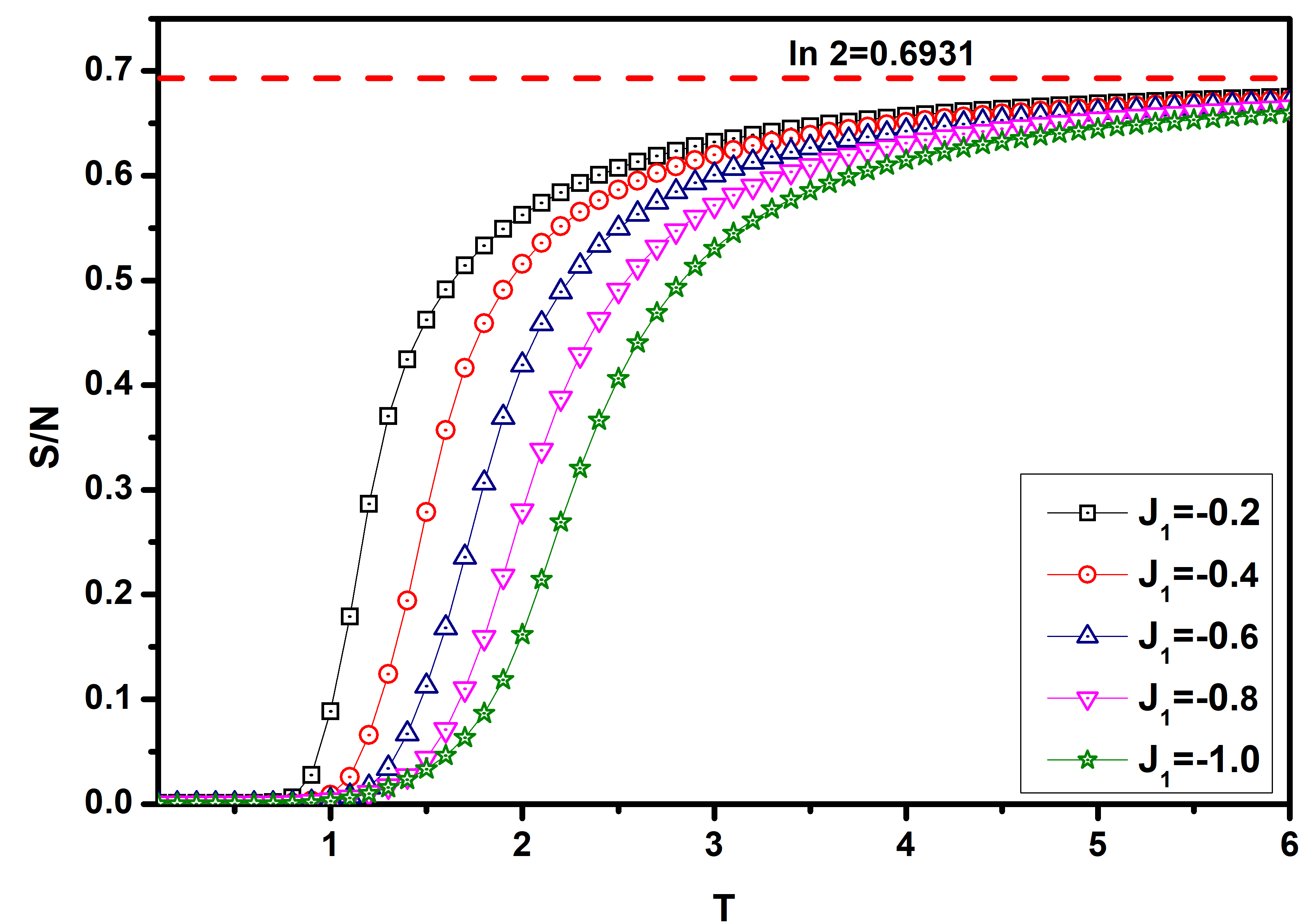}
		\caption{$ $}
		\label{f12d}
	\end{subfigure}
	\caption{Thermodynamic quantities for the SWINT without an applied magnetic field. (a) free energy and (b) entropy for A-type AFM; (c) free energy and (d) entropy for C-type AFM.}
	\label{f12}
\end{figure}

Free energy and entropy are calculated directly from the obtained DOS for A-type and C-type AFM orderings (Fig. \ref{f12}). As temperature increases, all free energy curves remain unchanged upto their transition temperatures. After that, the curves are gradually decreases and merged  (Figs. \ref{f12a} and \ref{f12c}). The entropy curves for both AFM orderings are given in Figs. \ref{f12b} and \ref{f12d}. The entropy value at low temperatures is close to zero as the number of accessible microstates is minimum. As the temperature increases, the number of accessible microstates increases, gradually raising the entropy curve. At high temperatures, the curves are merged and reach their saturation limit of the predicted magnetic entropy ($\ln2$).

\section*{Acknowledgement}
The author, A. Arul Anne Elden would like to thank T. Kiran for useful reading and valuable suggestions of the manuscript.

\end{document}